\documentclass{aa}  

\usepackage{graphicx}
\usepackage{txfonts}
\usepackage{natbib}
\usepackage{color}
\usepackage{multirow}
\usepackage{lscape}

\definecolor{red}{rgb}{1.,0.0,0.}

\usepackage{natbib,twoopt}

\usepackage[breaklinks=true, colorlinks=true, citecolor = blue]{hyperref} %% to avoid \citeads line fills
\bibpunct{(}{)}{;}{a}{}{,} %% natbib format for A&A and ApJ
\makeatletter
\newcommandtwoopt{\citeads}[3][][]{\href{http://adsabs.harvard.edu/abs/#3}%
        {\def\hyper@linkstart##1##2{}%
                \let\hyper@linkend\@empty\citealp[#1][#2]{#3}}}
\newcommandtwoopt{\citepads}[3][][]{\href{http://adsabs.harvard.edu/abs/#3}%
        {\def\hyper@linkstart##1##2{}%
                \let\hyper@linkend\@empty\citep[#1][#2]{#3}}}
\newcommandtwoopt{\citetads}[3][][]{\href{http://adsabs.harvard.edu/abs/#3}%
        {\def\hyper@linkstart##1##2{}%
                \let\hyper@linkend\@empty\citet[#1][#2]{#3}}}
\newcommandtwoopt{\citeyearads}[3][][]%
{\href{http://adsabs.harvard.edu/abs/#3}
        {\def\hyper@linkstart##1##2{}%
                \let\hyper@linkend\@empty\citeyear[#1][#2]{#3}}}
\makeatother

\usepackage{etoolbox}
\makeatletter
\patchcmd\@combinedblfloats{\box\@outputbox}{\unvbox\@outputbox}{}{%
        \errmessage{\noexpand\@combinedblfloats could not be patched}%
}%
\makeatother

\begin{document}

        \title{Fundamental properties of red-clump stars from long-baseline $H$-band interferometry\thanks{Based on observations made with ESO telescopes at the La Silla-Paranal observatory under programme IDs 092.D-0297, 094.D-0074 and 4100.L-0105}}
        \titlerunning{Fundamental properties of red-clump stars}
        %\subtitle{I. Overviewing the $\kappa$-mechanism}
        
        \author{A.~Gallenne\inst{1},
                G.~Pietrzy\'nski\inst{2,3},
                D.~Graczyk\inst{3,4,5},
                N.~Nardetto\inst{6},
                A.~M\'erand\inst{7},
                P.~Kervella\inst{8},
                W.~Gieren\inst{3,4},
                S.~Villanova\inst{3}
                R.~E.~Mennickent\inst{3},
                \and B.~Pilecki\inst{2}
                %P.~Kervella\inst{6, 7}
        }
        \authorrunning{Gallenne et al.}
        
        \institute{European Southern Observatory, Alonso de C\'ordova 3107, Casilla 19001, Santiago, Chile
                \and Nicolaus Copernicus Astronomical Centre, Polish Academy of Sciences,  Bartycka 18, 00-716 Warszawa, Poland
                \and Universidad de Concepci\'on, Departamento de Astronom\'ia, Casilla 160-C, Concepci\'on, Chile
                \and Millenium Institute of Astrophysics, Av. Vicu{\~n}a Mackenna 4860, Santiago, Chile
                \and Centrum Astronomiczne im. Miko\l{}aja Kopernika, PAN, Rabia\'nska 8, 87-100 Toru\'n, Poland
                \and Laboratoire Lagrange, UMR7293, Universit\'e de Nice Sophia-Antipolis, CNRS, Observatoire de la C\^ote d'Azur, Nice, France
                \and European Southern Observatory, Karl-Schwarzschild-Str. 2, 85748 Garching, Germany
                \and LESIA (UMR 8109), Observatoire de Paris, PSL, CNRS, UPMC, Univ. Paris-Diderot, 5 place Jules Janssen, 92195 Meudon, France
                %\and Unidad Mixta Internacional Franco-Chilena de Astronom\'{i}a (CNRS UMI 3386), Departamento de Astronom\'{i}a, Universidad de Chile, Camino El Observatorio 1515, Las Condes, Santiago, Chile
        }
        
        %\date{Received September 15, 1996; accepted March 16, 1997}
        
        % \abstract{}{}{}{}{} 
        % 5 {} token are mandatory
        
        \abstract
        %       {\red{TO DO}}
        %       {\red{TO DO}}
        %       {\red{TO DO}}
        %       {\red{TO DO}}
        {Observations of 48 red-clump stars were obtained in the $H$ band with the PIONIER instrument installed at the Very Large Telescope Interferometer. Limb-darkened angular diameters were measured by fitting radial intensity profile $I(r)$ to square visibility measurements. Half the angular diameters determined have formal errors better than 1.2\,\%, while the overall accuracy is better than 2.7\,\%. Average stellar atmospheric parameters (effective temperatures, metallicities and surface gravities) were determined from new spectroscopic observations and literature data and combined with precise Gaia parallaxes to derive a set of fundamental stellar properties. These intrinsic parameters were then fitted to existing isochrone models to infer masses and ages of the stars. The added value from interferometry imposes a better and independent constraint on the $R-\mathrm{T_{eff}}$ plane. Our derived values are consistent with previous works, although there is a strong scatter in age between various models. This shows that atmospheric parameters, mainly metallicities and surface gravities, still suffer from a non-accurate determination, limiting constraints on input physics and parameters of stellar evolution models.}
        
        
        \maketitle
        %
        %-------------------------------------------------------------------
        
        \section{Introduction}
        
        Direct stellar angular diameter measurements are a valuable observable with which to determine a star's fundamental properties, particularly the linear radius and absolute luminosity through the combination of Gaia parallaxes. These properties, together with the mass, are particularly necessary to constrain stellar structure and evolution models. Usually, spectroscopy is used to determine the effective temperatures, surface gravities, and metallicities of a star which, combined with a known distance, provide the stellar radius and luminosity. This is then fitted to evolutionary tracks to yield the stellar mass. However, the spectroscopic atmospheric parameters strongly depend on the atmosphere models used, and do not provide estimates accurate enough to well constrain stellar evolution models. Various models exist in the literature, using different input physics and parameters (e.g. the helium content, mixing length parameter, amongst others), which can only be constrain with more precise measurements \citep[see e.g.][]{Gallenne_2016_02_0,Valle_2017_04_0}. Additional accurate parameters such as angular diameters provide independent constraints on the linear radii and luminosities.
        
        The high angular resolution obtained from long-baseline interferometry (LBI) enables us to spatially resolve the photospheric disks of the apparent biggest stars (typically a diameter $> 0.5$\,mas, i.e. $\sim 10\,\mathrm{R_\odot}$ at 100\,pc). LBI can provide very accurate angular size measurements, as already demonstrated \citep[see e.g.][]{Nordgren_1999_12_0,Mozurkewich_2003_11_0,Kervella_2004_03_0,Baines_2010_02_0,Boyajian_2012_02_0,Gallenne_2012_03_0}, and provide a valuable constraint on the input physics of theoretical stellar models. 
        
        In this paper, we present the measurements of the angular diameter of 48 F and G-type red-clump giant stars observed with the Very Large Telescope Interferometer (VLTI). The purpose is to accurately determine the absolute properties of such kind of stars through additional observable constraints. Details on the observations and data reduction are presented in Sect.~\ref{section__observations_and_data_reduction}, including additional spectroscopic observations. Sect.~\ref{section_limb_darkened_angular_diameters} is dedicated to the determination of limb-darkened angular diameters {using atmospheric models}. Stellar properties and derived masses and ages are presented in Sect.~\ref{section__stellar_properties}, and we summarize in Sect.~\ref{section_conclusion}.
        
        \section{Observations and data reduction}
        \label{section__observations_and_data_reduction}
        
        \subsection{Selected targets}
        
        All our red-clump stars were selected from \citet{Laney_2012_01_0}, for which uniform and accurate near-infrared magnitudes have been obtained ($\sim 0.005$\,mag). The initial goal of these interferometric observations was to measure their angular diameter to calibrate the surface brightness-colour (SBC) relation for late-type stars. These diameters were then combined with high-quality and homogeneous $V$- and $K$-band photometry \citep{Mermilliod_1997_08_0,Laney_2012_01_0} following the relation $S_\mathrm{V} = V_0 +5~\log{\theta_\mathrm{LD}} = a~(V - K)_0 + b$. This new accurate calibration of $a$ and $b$, specific for these stars, is published in \citet{Pietrzynski_2018__0}, and enables the determination of angular diameters at a 0.8\,\% accuracy level (r.m.s. of the SBC relation of 0.018\,mag). We then used this relation for late-type eclipsing systems to measure the most accurate distance of the Large Magellanic Cloud to 1\,\% \citep{Pietrzynski_2018__0}. This provides the best reference point ever obtained for the cosmic distance scale. Furthermore, accurate parallaxes for these red-clump stars can be found in the Gaia DR2 \citep{Gaia-Collaboration_2018_04_0,Gaia-Collaboration_2016_11_0}, and combining them with angular diameter measurements provides unbiased informations on their intrinsic fundamental parameters, such as the linear radius and luminosity.
        
        Our selection criteria in the dataset of \citet{Laney_2012_01_0} were to choose:
        \begin{itemize}
                \item targets with declination $< 20^\circ$ in order to be observable from the VLTI,
                \item expected angular diameters $> 0.8$\,mas to be sufficiently resolved by the longest available VLTI baseline,
                \item not flagged as binary in the \texttt{Simbad} database
        \end{itemize}
        
        Also taking also into account the observability of the targets for visitor mode observations, we finally ended up with a total of 48 targets.

        \subsection{VLTI/PIONIER interferometric observations}
        
        \begin{figure}
                \centering
                \resizebox{\hsize}{!}{\includegraphics[width = \linewidth]{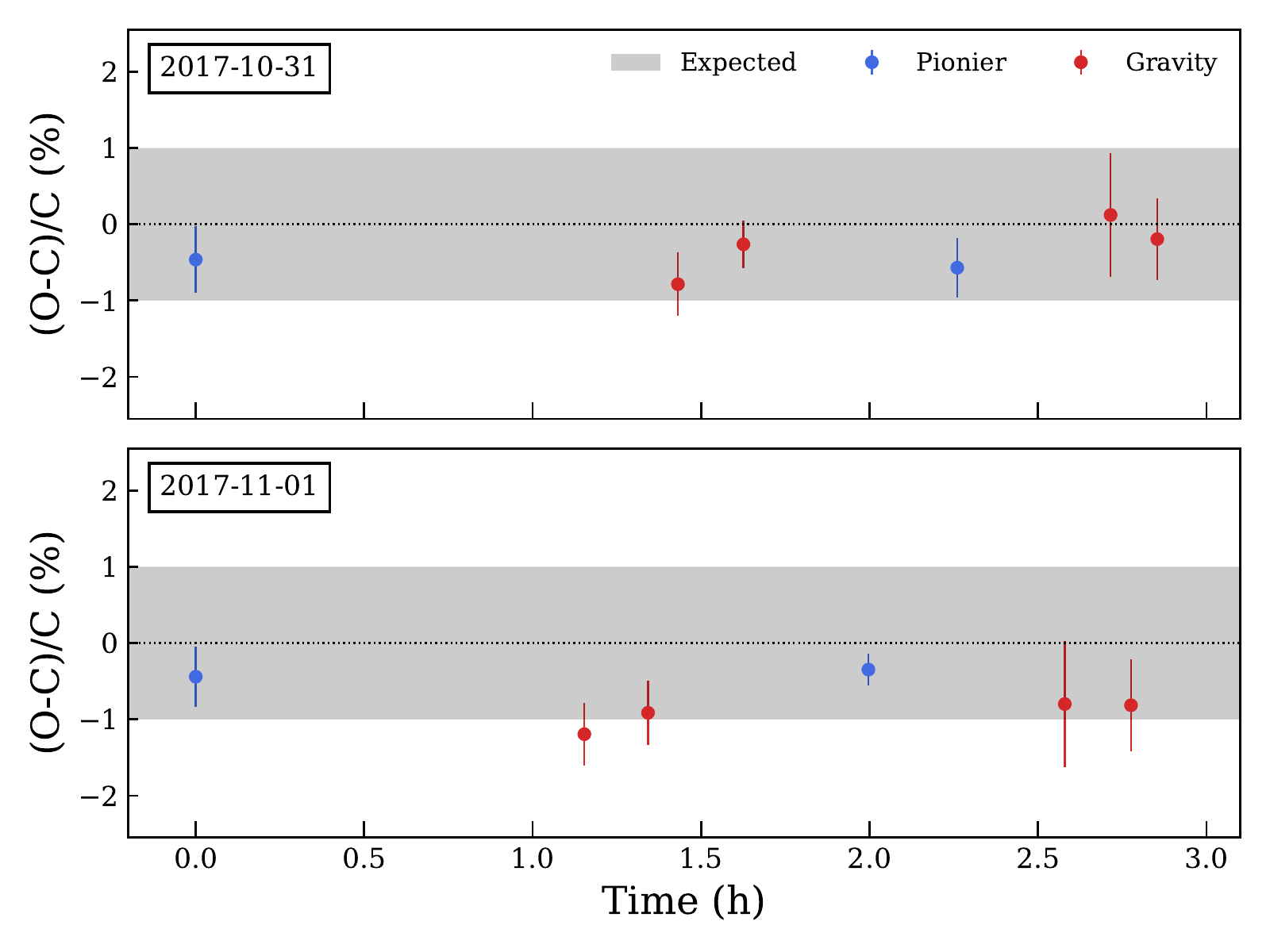}}
                \caption{Relative difference of the observed and calculated projected separations.}
                \label{figure_wave}
        \end{figure}
        
        We used the Very Large Telescope Interferometer \citep[VLTI ;][]{Haguenauer_2010_07_0} with the four-telescope combiner PIONIER \citep[Precision Integrated Optics Near-infrared Imaging ExpeRiment,][]{Le-Bouquin_2011_11_0} to measure squared visibilities ($V^2$) and closure phases ($CP$) of our red-clump stars. PIONIER combines the light coming from four telescopes in the $H$ band, either in a broad band mode or with a low spectral resolution, where the light is dispersed across a few spectral channel. Before Dec.~2014, the fringe dispersion was possible across three or seven channels, then PIONIER was upgraded with a new detector and a new GRISM dispersion mode with six spectral channels.
        
        Our observations were carried out from 2013 to 2015 using the 1.8\,m Auxiliary Telescopes with the configurations A1-G1-J3-K0 and A1-G1-I1-K0, providing six projected baselines ranging from 45 to 140\,m. PIONIER was set up in dispersed mode for all targets, that is, the fringes are dispersed into three, seven, or six spectral channels.
        
        We monitored the interferometric transfer function with the standard procedure which consists of interleaving the science target by reference stars. The calibrators were selected using the \textit{SearchCal} software\footnote{Available at \url{http://www.jmmc.fr/searchcal.}} \citep{Bonneau_2006_09_0,Bonneau_2011_11_0} provided by the Jean-Marie Mariotti Center\footnote{\url{http://www.jmmc.fr}} (JMMC), and are listed in Table~\ref{table__journal}, together with the journal of the observations.
The data have been reduced with the \textit{pndrs} package described in \citet{Le-Bouquin_2011_11_0}. The main procedure is to compute squared visibilities and triple products for each baseline and spectral channel, and to correct for photon and readout noises.
        
\longtab{
	\begin{landscape}
		\begin{longtable}{cccccccccc}
			\caption{\label{table__journal} Log of the interferometric observations, together with some stellar information.} \\
			\hline\hline
			Star  & $K$ &  $H$  &  $\pi$    & $E(B-V)$ &  Date  &   Baselines  &   Sp.  &   Calibrator\tablefootmark{b}  & Bin.   \\
			& (mag) & (mag) & (mas)    &                            &               &                                        &    channels &                         HD                                                                       &          \\
			\hline
			\object{HD~360}   & 3.653  &   $3.757$   & $8.97\pm0.13$   & 0.009 &  2014-10-05  &  A1-G1-J3-K0  &  3   &    \object{HD~1588}, 6482  & -     \\                
			\object{HD~3750}  & 3.485 &   $3.612$   & $10.26\pm0.11$   & 0.002 &  2013-12-31  &  A1-G1-J3-K0  &  7   &   \object{HD~3145}, \object{HD~224821}, \object{HD~902}   & -   \\
			%                                                       &                         &                                       &                               &                                 &                                               &                                               &               &               \object{HD~902} \\
			\object{HD~4211}  & 3.295 &     $3.426$   & $9.63\pm0.40$\tablefootmark{a}   & 0.004  & 2014-10-05  &  A1-G1-J3-K0  &  3   &    \object{HD~1434}, 902, \object{HD~3145}   & -    \\
			%                                                       &                       &                                       &                               &                                 &                                               &                                               &               &               \object{HD~3145}        \\
			\object{HD~5722}    & 3.381 &    $3.496$   & $9.79\pm0.15$   & 0.010 & 2014-01-01  &  A1-G1-J3-K0  &  7   &     \object{HD~3975}, \object{HD~3909}, \object{HD~6482} & -    \\
			%                                                       &                               &                                       &                               &                                 &                                               &                                                       &               &               \object{HD~6482}        \\
			\object{HD~8651}  &  3.019  &     $3.142$   & $12.82\pm0.15$   & 0.002 & 2013-10-14  &  A1-G1-J3-K0  &  7   &   HD~10216, HD~11050  & -       \\
			\object{HD~9362}  & 1.638  &     $1.748$   & $22.95\pm0.19$\tablefootmark{a}   & 0.000 & 2014-10-08  &  A1-G1-J3-K0  &  3   &     \object{HD~9293}, \object{HD~8963}  & -    \\
			\object{HD~10142}  &  3.557  &   $3.679$   & $9.80\pm0.13$   & 0.007     &  2013-12-31  &  A1-G1-J3-K0  &  7   &  \object{HD~10216}, \object{HD~11050}, \object{HD~11643}  &  -     \\
			%                                                       &                               &                                       &                               &                                 &                                               &                                                       &               &               \object{HD~11643}       \\
			&                               &                                        &                       &                               &  2014-10-05  &  A1-G1-J3-K0  &  3   &    \object{HD~9742}, \object{HD~8901}       \\
			&                               &                               &                         &                               &  2014-10-31  &  A1-G1-J3-K0  &  3   &    \object{HD~14509}, \object{HD~14832}       \\
			\object{HD~11977}  &  2.486  & $2.594$   & $14.72\pm0.18$   & 0.002&  2013-10-14  &  A1-G1-J3-K0  &  7   &   \object{HD~13668}, \object{HD~18423}, \object{HD~18959}    & -     \\
			&                               &                               &                                 &                       &  2014-10-06  &  A1-G1-J3-K0  &  3   &     \object{HD~21208}, HD~12851, \object{HD~18185}   \\
			%                                                       &                               &                                       &                                       &                         &                                               &                                                       &               &               \object{HD~18185}       \\
			&                               &                               &                                 &                       &  2014-12-19  &  A1-G1-I1-K0  &  6   &      \object{HD~18959}     \\
			\object{HD~12438}  &  3.176  &   $3.281$   & $11.08\pm0.29$\tablefootmark{a}   & 0.004 &  2013-10-14  &  A1-G1-J3-K0  &  7   &    HD~11050, \object{HD~15958}  & -     \\
			\object{HD~13468}  &  3.666  &   $3.783$   & $8.52\pm0.12$   & 0.009     &  2014-10-05  &  A1-G1-J3-K0  &  3   &     HD~19121, HD~14129  & -    \\
			\object{HD~15220}  &  3.199  &   $3.342$   & $11.89\pm0.42$\tablefootmark{a}   & 0.007 &  2014-01-02  &  A1-G1-J3-K0  &  7   &    \object{HD~15996}, \object{HD~13692}, \object{HD~18290}  & -      \\
			%                                                       &                               &                                       &                                       &                        &                                               &                                                       &               &               \object{HD~18290}       \\
			\object{HD~15248}  &  3.553  &   $3.669$   & $9.28\pm0.11$   & 0.010 &  2014-10-05  &  A1-G1-J3-K0  &  3   &     \object{HD~19755}, \object{HD~12851}, \object{HD~18696}  & -    \\
			%                                                       &                               &                                       &                                       &                         &                                               &                                                       &               &               \object{HD~18696}       \\
			\object{HD~15779}  &   3.067  &  $3.186$   & $12.06\pm0.14$   & 0.006 &  2014-01-01  &  A1-G1-J3-K0  &  7   &  \object{HD~14129}, \object{HD~19121}, \object{HD~20791}  & -     \\
			%                                                               &                               &                                       &                               &                         &                                               &                                                       &               &               \object{HD~20791}       \\
			&                               &                               &                                  &                       &  2014-10-05  &  A1-G1-J3-K0  &  3   &   \object{HD~14690}, \object{HD~13819}        \\
			\object{HD~16815}  &  1.706  &   $1.820$   & $21.65\pm0.18$\tablefootmark{a}   & 0.000 &  2014-10-08  &  A1-G1-J3-K0  &  3   &   \object{HD~13666}, \object{HD~15875}  & -      \\
			\object{HD~17652}  &  2.139  &   $2.256$   & $18.89\pm0.26$\tablefootmark{a}   & 0.001 &  2014-12-19  &  A1-G1-I1-K0  &  6   &    \object{HD~15471}, HD~20176  & -      \\
			\object{HD~17824}  &  2.668  &   $2.781$   & $16.92\pm0.30$   & 0.002 &  2014-10-06  &  A1-G1-J3-K0  &  3   &     \object{HD~18071}, \object{HD~20520}  & -    \\
			&                               &                               &                                 &                        &  2014-11-17  &  A1-G1-I1-K0  &  3   &     HD~13692, HD~18071      \\
			&                               &                               &                                          &                      &  2014-12-19  &  A1-G1-I1-K0  &  6   &      \object{HD~14728}, HD~13692     \\
			\object{HD~18784}  &  3.353  &  $3.456$   & $10.49\pm0.13$   & 0.014  &  2014-10-05  &  A1-G1-J3-K0  &  3   &     HD~14129, HD~19121 & -     \\
			\object{HD~23319}  &  1.995  &   $2.141$   & $19.22\pm0.25$   & 0.001 &  2014-10-08  &  A1-G1-J3-K0  &  3   &  \object{HD~21149}, \object{HD~26934}    & -     \\
			&                               &                               &                                 &                        &  2015-01-16  &  A1-G1-I1-K0  &  6   &   HD~21149, HD~26934, \object{HD~24267}      \\
			%                                                       &                                 &                                       &                                       &                         &                                               &                                                       &               &               \object{HD~24267}       \\
			\object{HD~23526}  &  3.634  &   $3.744$   & $8.85\pm0.14$   & 0.017 &  2014-10-05  &  A1-G1-J3-K0  &  3   &   HD~20791, HD~19121 & -       \\
			&                               &                               &                                 &                        &  2014-10-31  &  A1-G1-J3-K0  &  3   &    HD~20791, \object{HD~28322}       \\
			&                               &                               &                                 &                               &  2014-12-17  &  A1-G1-I1-K0  &  6   &    HD~20791, HD~28322       \\
			\object{HD~23940}  &  3.229  &   $3.344$   & $12.13\pm0.30$\tablefootmark{a}   & 0.002 &  2013-12-31  &  A1-G1-J3-K0  &  7   &   \object{HD~20176}, \object{HD~22826}   & -      \\
			\object{HD~26464}  &  3.341  &   $3.449$   & $10.04\pm0.13$   & 0. 008 &  2014-10-05  &  A1-G1-J3-K0  &  3   &   \object{HD~27179}, \object{HD~28947}   & x     \\
			\object{HD~30814}  &   2.791  &  $2.898$   & $13.62\pm0.20$   & 0.006 &  2014-01-02  &  A1-G1-J3-K0  &  7   &  \object{HD~28625}, \object{HD~31887}, \object{HD~32613}  & -    \\
			%                                                       &                               &                                       &                                       &                                 &                                               &                                                       &               &               \object{HD~32613}       \\
			\object{HD~35369}  &  1.925  &   $2.032$   & $18.75\pm0.33$   & 0.000 &  2014-10-08  &  A1-G1-J3-K0  &  3   &    \object{HD~32707}, \object{HD~36134}  & x    \\
			&                               &                               &                                 &                       &  2014-11-17  &  A1-G1-I1-K0  &  3   &     \object{HD~38054}, HD~40605      \\
			\object{HD~36874}  &  3.242  &   $3.364$   & $11.24\pm0.12$   & 0.002 &  2014-10-05  &  A1-G1-J3-K0  &  3   &   \object{HD~37377}, \object{HD~38885}  & -       \\
			\object{HD~39523}  &  2.036  &   $2.160$   & $18.59\pm0.26$  & 0.001 & 2014-10-08  &  A1-G1-J3-K0  &  3   &   HD~26934, \object{HD~42026}, \object{HD~37462}    & -     \\
			&                               &                               &                                                                 &                       &  2014-10-31  &  A1-G1-J3-K0  &  3   &  HD~42026, HD~37462        \\
			\object{HD~39640}  &  2.921  &    $3.040$   & $12.54\pm0.15$   & 0.006 & 2014-10-05  &  A1-G1-J3-K0  &  3   &   \object{HD~37877}, HD~34587  & -      \\
			&                               &                               &                                 &                       &  2014-10-31  &  A1-G1-J3-K0  &  3   &    \object{HD~47001}, \object{HD~42168}       \\
			&                               &                               &                                 &                       &  2014-11-17  &  A1-G1-I1-K0  &  3   &    HD~37877, HD~38054       \\
			&                               &                               &                                 &                       &  2015-01-09  &  A1-G1-I1-K0  &  6   &    HD~37877, HD~34587       \\
			\object{HD~39910}  &  3.315  &    $3.451$   & $10.43\pm0.13$   & 0.015 & 2014-10-05  &  A1-G1-J3-K0  &  3   &   HD~34137, \object{HD~40605}  & -      \\
			\object{HD~40020}  &  3.419  &   $3.543$   & $9.69\pm0.14$   & 0.013 & 2014-01-02  &  A1-G1-J3-K0  &  7   &  HD~34137   & -      \\
			\object{HD~43899}  &  3.004  &   $3.134$   & $11.66\pm0.11$   & 0.010 & 2013-10-14  &  A1-G1-J3-K0  &  7   &   \object{HD~52574}, \object{HD~44956}   & -    \\
			\object{HD~45415}  &  3.188  &  $3.306$   & $11.03\pm0.18$   & 0.015 &  2015-01-16  &  A1-G1-I1-K0  &  6   &   \object{HD~44769}, \object{HD~55185}   & x     \\
			\object{HD~46116}  &  3.103  &   $3.206$   & $12.08\pm0.18$\tablefootmark{a}   & 0.009 & 2014-10-05  &  A1-G1-J3-K0  &  3   &   \object{HD~51801}, \object{HD~39810}    & -    \\
			\object{HD~53629}  &  3.410  &   $3.556$   & $9.69\pm0.11$   & 0.017 &  2014-01-02  &  A1-G1-J3-K0  &  7   &    \object{HD~51546}, \object{HD~57911}   & -    \\
			\object{HD~54131}  &  3.114  &    $3.222$   & $11.24\pm0.15$   & 0.012 & 2015-01-16  &  A1-G1-I1-K0  &  6   &     \object{HD~56537}  & -    \\
			\object{HD~56160}  &  2.823  &    $2.976$   & $12.27\pm0.13$   & 0.010 & 2014-10-31  &  A1-G1-J3-K0  &  3   &     \object{HD~54257}, HD~57911  & -    \\
			\object{HD~60060}  &  3.545  &    $3.662$   & $9.33\pm0.11$   & 0.018 & 2014-10-31  &  A1-G1-J3-K0  &  3   &    \object{HD~52603}, \object{HD~62897}  & -     \\
			&                               &                               &                                 &                       &  2014-12-17  &  A1-G1-J3-K0  &  3   &     HD~53840, HD~52603      \\
			&                               &                               &                                 &                       &  2014-12-19  &  A1-G1-I1-K0  &  6   &     \object{HD~53840}, HD~62897      \\
			\object{HD~60341}  &  3.126  &    $3.264$   & $11.10\pm0.13$   & 0.010  & 2014-01-02  &  A1-G1-J3-K0  &  7   &  \object{HD~56110}, \object{HD~57820}  & -       \\
			\object{HD~62412}  &  3.412  &    $3.530$   & $9.88\pm0.12$   & 0.013 & 2014-01-02  &  A1-G1-J3-K0  &  7   &   HD~57911, \object{HD~70097}  & x      \\
			&                               &                               &                                 &                       &  2015-01-09  &  A1-G1-I1-K0  &  6   &    HD~57911, HD~57820, HD~51546       \\
			\object{HD~62713}  &  2.654  &     $2.782$   & $15.03\pm0.18$\tablefootmark{a}   & 0.005 & 2014-10-06  &  A1-G1-J3-K0  &  3   &  \object{HD~51682}, \object{HD~68512}   & -      \\
			\object{HD~68312}  &  3.279  &     $3.390$   & $11.65\pm0.17$   & 0.011 & 2014-01-02  &  A1-G1-J3-K0  &  7   &  \object{HD~71465}, \object{HD~70136}   & -      \\
			&                               &                               &                                 &                       &  2015-02-17  &  A1-G1-J3-K0  &  6   &    HD~70409, HD~70136       \\
			\object{HD~74622}  &  3.532  &     $3.672$   & $10.32\pm0.10$   & 0.013 & 2014-10-05  &  A1-G1-J3-K0  &  3   &   \object{HD~81720}   & -    \\
			\object{HD~75916}  &  3.516  &     $3.653$   & $9.41\pm0.11$   & 0.008 & 2015-02-17  &  A1-G1-J3-K0  &  6   &    \object{HD~71231}, \object{HD~70409}  & -     \\
			\object{HD~176704}&  2.956 & $3.107$ & $12.66\pm0.29$\tablefootmark{a} & 0.007 &  2014-10-05  &  A1-G1-J3-K0  & 3 & \object{HD~176752}, \object{HD~171960}, \object{HD~174774} &  -     \\
			%                                                               &                               &                                       &                               &                         &                                               &                                                       &               &               \object{HD~174774}      \\
			\object{HD~177873}  &   2.137  &  $2.243$ & $18.73\pm0.33$ & 0.002        &  2014-06-25  &  A1-G1-J3-K0 &  3 & \object{HD~178272}, \object{HD~181110}, \object{HD~184349} & x      \\
			%                                                       &                               &                                       &                               &                                 &                                               &                                                       &               &               \object{HD~184349}      \\
			&                       &                                       &                                 &                       &  2014-10-05  &  A1-G1-J3-K0  &  3   &      HD~184349     \\
			\object{HD~188887}  &  2.621 & $2.763$  & $13.85\pm0.13$ & 0 &  2014-06-25  &  A1-G1-J3-K0  &  3   &  \object{HD~181019}, \object{HD~207229}, \object{HD~210563}  & x       \\
			%                                                       &                                       &                               &                                         &                                               &                                                       &               &               \object{HD~210563}      \\
			\object{HD~191584}  &  3.512  &   $3.664$   & $9.88\pm0.12$   & 0.009 &  2014-10-05  &  A1-G1-J3-K0  &  3   &   \object{HD~189563}, \object{HD~190057}, \object{HD~195659} & -       \\
			%                                                       &                               &                                       &                               &                                 &                                               &                                                       &               &               \object{HD~195659}      \\
			\object{HD~204381}  &  2.426  &   $2.537$   & $19.06\pm0.29$\tablefootmark{a}   & 0. &  2014-06-25  &  A1-G1-J3-K0  &  3   &   \object{HD~206146}, \object{HD~204609}  & -      \\
			\object{HD~219784}  & 1.886  &  $2.019$   & $19.61\pm0.35$   & 0.001 &  2014-10-05  &  A1-G1-J3-K0  &  3   &    \object{HD~221370}, \object{HD~214465}  & -     \\
			\object{HD~220572}  &  3.224  &   $3.350$   & $11.01\pm0.13$ & 0.003 &  2014-10-05 &  A1-G1-J3-K0  &  3 & \object{HD~220330}, \object{HD~220790}, \object{HD~215905}  & -     \\
			%                                                       &                               &                                       &                               &                                 &                                               &                                                       &               &               \object{HD~215905}      \\
			\hline
		\end{longtable}
		\tablefoot{$K$- and $H$-band magnitudes are from \citet{Laney_2012_01_0} with uncertainties of 0.005\,mag, and parallaxes from the Gaia DR2. $E(B-V)$ were determined as explained in \citet{Suchomska_2015_07_0}.\\
			\tablefoottext{a}{From the Hipparcos catalogue \citep{van-Leeuwen_2007_11_0}.}\\
			\tablefoottext{b}{$H$-band uniform disk diameter of the calibrators:
				HD3145=$0.852\pm0.012$\,mas;HD224821=$0.929\pm0.013$\,mas;HD902=$0.967\pm0.013$\,mas;
				HD11050=$0.710\pm0.009$\,mas;HD10216=$0.834\pm0.011$\,mas;HD11643=$0.949\pm0.013$\,mas;
				HD20176=$0.897\pm0.012$\,mas;HD22826=$0.742\pm0.010$\,mas;
				HD13668=$0.784\pm0.010$\,mas;HD18423=$1.092\pm0.014$\,mas;
				HD34587=$0.884\pm0.012$\,mas;HD47001=$0.867\pm0.011$\,mas;
				HD15958=$0.864\pm0.012$\,mas;
				HD52574=$0.675\pm0.009$\,mas;HD44956=$0.655\pm0.008$\,mas;
				HD3975=$0.957\pm0.013$\,mas;HD3909=$0.861\pm0.012$\,mas;HD6482=$0.836\pm0.012$\,mas;
				HD14129=$1.006\pm0.014$\,mas;HD19121=$0.890\pm0.012$\,mas;HD20791=$0.883\pm0.012$\,mas;
				HD15996=$0.870\pm0.011$\,mas;HD13692=$0.923\pm0.012$\,mas;HD18290=$0.779\pm0.011$\,mas;
				HD15471=$0.911\pm0.013$\,mas;
				HD28625=$0.754\pm0.009$\,mas;HD31887=$0.742\pm0.010$\,mas;HD32613=$0.864\pm0.012$\,mas;
				HD34137=$0.798\pm0.011$\,mas;HD81720=$0.922\pm0.013$\,mas;
				HD56110=$0.817\pm0.011$\,mas;HD57820=$0.959\pm0.013$\,mas;
				HD71465=$1.043\pm0.011$\,mas;HD70136=$0.923\pm0.013$\,mas;
				HD51546=$0.865\pm0.012$\,mas;HD57911=$0.877\pm0.012$\,mas;
				HD49001=$1.055\pm0.014$\,mas;HD40605=$0.978\pm0.013$\,mas;
				HD70097=$0.916\pm0.013$\,mas;HD37877=$0.951\pm0.013$\,mas;
				HD178272=$1.005\pm0.014$\,mas;HD181110=$0.945\pm0.013$\,mas;HD184349=$1.067\pm0.015$\,mas;
				HD181019=$1.030\pm0.014$\,mas;HD207229=$1.001\pm0.013$\,mas;HD210563=$0.968\pm0.013$\,mas;
				HD206146=$1.126\pm0.015$\,mas;HD204609=$1.143\pm0.016$\,mas;
				HD176752=$1.198\pm0.017$\,mas;HD171960=$1.121\pm0.016$\,mas;HD174774=$1.103\pm0.015$\,mas;
				HD189563=$0.915\pm0.013$\,mas;HD190057=$1.037\pm0.014$\,mas;HD195659=$0.929\pm0.012$\,mas;
				HD220330=$0.953\pm0.013$\,mas;HD220790=$1.007\pm0.013$\,mas;HD215905=$0.995\pm0.013$\,mas;
				HD1434=$0.941\pm0.013$\,mas;HD1588=$0.892\pm0.012$\,mas;
				HD221370=$0.906\pm0.013$\,mas;HD214465=$1.193\pm0.016$\,mas;
				HD9742=$0.942\pm0.013$\,mas;HD8901=$0.968\pm0.013$\,mas;
				HD19755=$0.840\pm0.012$\,mas;HD12851=$0.920\pm0.013$\,mas;HD18696=$0.950\pm0.012$\,mas;
				HD10164=$1.061\pm0.014$\,mas;HD6903=$0.628\pm0.044$\,mas;
				HD14690=$0.475\pm0.033$\,mas;HD13819=$0.541\pm0.038$\,mas;
				HD37377=$0.890\pm0.011$\,mas;HD38885=$1.000\pm0.013$\,mas;
				HD27179=$0.919\pm0.013$\,mas;HD28947=$0.722\pm0.008$\,mas;
				HD51801=$0.826\pm0.011$\,mas;HD39810=$0.690\pm0.008$\,mas;
				HD18185=$1.141\pm0.015$\,mas;HD81502=$1.230\pm0.016$\,mas;
				HD18071=$1.024\pm0.014$\,mas;HD20520=$1.267\pm0.017$\,mas;
				HD51682=$1.034\pm0.014$\,mas;HD68512=$1.227\pm0.017$\,mas;
				HD9293=$1.232\pm0.017$\,mas;HD8963=$1.076\pm0.015$\,mas;
				HD13666=$1.007\pm0.014$\,mas;HD15875=$0.915\pm0.012$\,mas;
				HD32707=$1.169\pm0.016$\,mas;HD36134=$1.164\pm0.016$\,mas;
				HD21149=$1.056\pm0.014$\,mas;HD26934=$0.984\pm0.013$\,mas;
				HD14509=$0.866\pm0.012$\,mas;HD14832=$0.732\pm0.009$\,mas;
				HD52603=$0.889\pm0.012$\,mas;HD62897=$0.834\pm0.011$\,mas;
				HD54257=$0.839\pm0.012$\,mas;HD28322=$0.819\pm0.011$\,mas;
				HD42168=$0.806\pm0.011$\,mas;HD38054=$1.266\pm0.017$\,mas;
				HD42026=$1.098\pm0.014$\,mas;HD37462=$1.263\pm0.017$\,mas;
				HD18959=$1.140\pm0.016$\,mas;HD14728=$1.171\pm0.016$\,mas;
				HD53840=$0.830\pm0.010$\,mas;HD56537=$0.567\pm0.040$\,mas;
				HD44769=$0.583\pm0.041$\,mas;HD55185=$0.533\pm0.037$\,mas;
				HD24267=$1.064\pm0.014$\,mas;
				HD71231=$0.896\pm0.012$\,mas;HD70409=$0.842\pm0.012$\,mas;
			}
		}
	\end{landscape}
}
        
        \begin{figure*}
                \centering
                \resizebox{\hsize}{!}{\includegraphics[width = \linewidth]{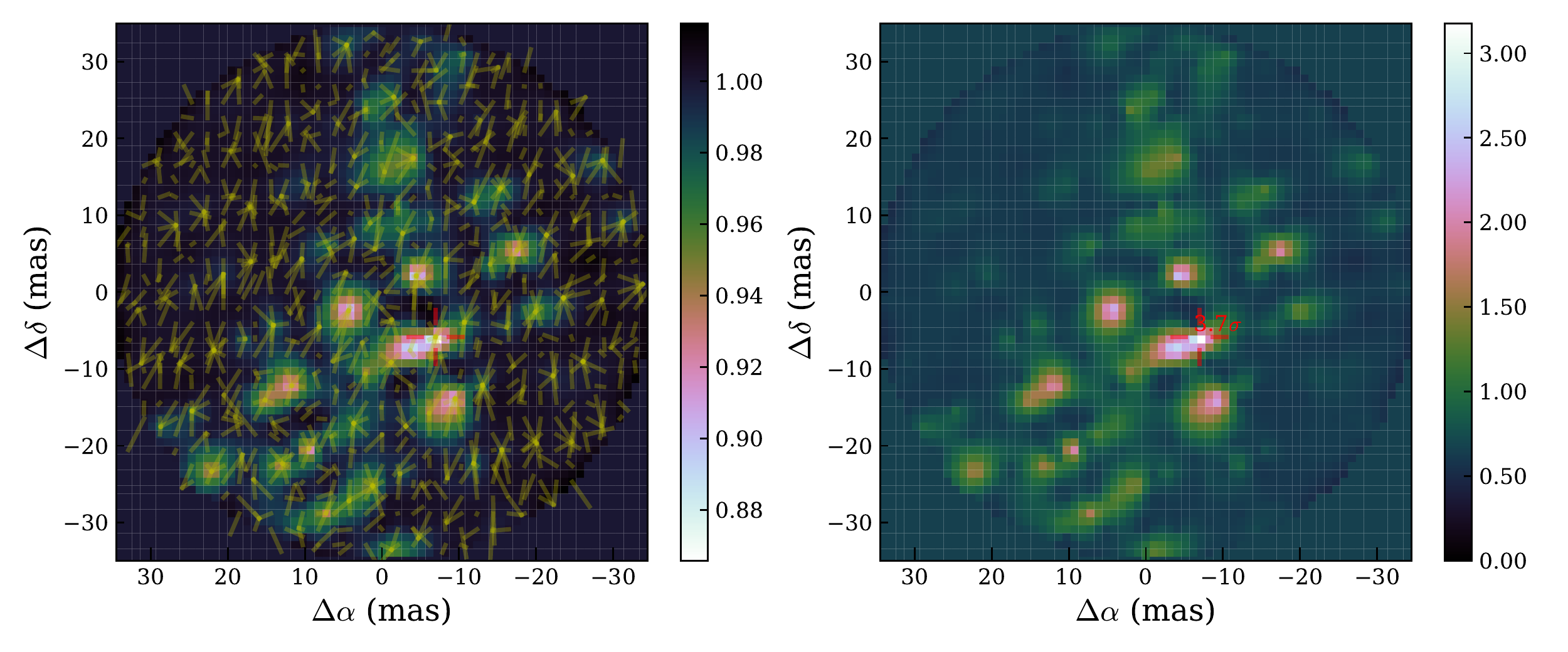}}
                \caption{$\chi^2$ map of the local minima (left) and detection level map (right) of hd45415. The yellow lines represent the convergence from the starting points to the final fitted position. The maps were reinterpolated in a regular grid for clarity.}
                \label{figure_chi2map}
        \end{figure*}
        
        \subsection{Wavelength calibration}
        
        There are several sources of systematic in interferometry, but most of them are known and reduced to less than 0.1\,\%. For the interferometric observable, the visibility, which is a function of the spatial frequency $B/\lambda$, the main systematic is the accuracy on the wavelength calibration. The VLTI baseline lengths during the observations are known at an accuracy of better then 0.01\,\% thanks to a metrology laser, but the accuracy of the spectral calibration is instrument-dependent. In the case of PIONIER, this calibration is linked to scanning piezos used for the optical path delay modulation.
        
        We performed during two hours several spectral calibrations on a cloudy night, with PIONIER in order to check the stability and repeatability (of the wavelength of each spectral channel). Over the two hours duration of the test, without moving the  optics, the repeatability was found to be precise to $\sim 0.02$\,\%. The comparison with the spectral calibration taken at the beginning of night (6h before) is precise to $\sim 0.06$\,\%, while the comparison with the end of the night calibration, after moving the optics, gives an immediate repeatability precise to $\sim 0.02$\,\%. However, there is an additional systematic error because the accuracy is limited by the calibration of the scanning piezo, which is usually assumed to be accurate to about 1\,\%. To better quantify this, we performed specific observations with GRAVITY, the second generation interferometric instrument of the VLTI \citep{Eisenhauer_2011_03_0}, which has a dedicated internal reference laser source allowing a wavelength accuracy better than 0.02\,\%. GRAVITY can therefore be used to cross-calibrate PIONIER through observations of a same target.
        
        Our calibration observations consisted of observing the very well known binary star \object{TZ~For}, for which the orbital parameters were derived with exquisite accuracy by \citet{Gallenne_2016_02_0}. The observations were executed over two half nights on 31 October and 1 November 2017, alternating between both GRAVITY (hereafter G) and PIONIER (hereafter P). For each instrument we used the usual interferometric observing method by interleaving the TZ~For observations by calibrator stars in order to monitor the transfer function of each instrument. The observing sequence with the instruments for each half night was P-G-P-G (we will call a sequence P-G a dataset). Data were reduced with the corresponding instrument pipeline, and the relative astrometric position of the secondary for each observation determined using the \texttt{CANDID} tool (see Sect.~\ref{section__checking_for_binarity}). In Fig.~\ref{figure_wave} we plot the relative difference between our observed projected separations for each instrument and the ones calculated from the orbital solutions of \citet{Gallenne_2016_02_0}. We see a very good agreement in results between the two instruments and with the calculated position. For the first half night, we measured a relative difference of 0.22\,\% for the first data set (first P-G), and 0.28\,\% for the second one. For the first half night we found 0.29\,\%. For the second half night, we determined a relative difference of 0.31\,\% for the first data set, and 0.22\,\% for the second one. For the whole half night we also found 0.29\,\%. Combining all the data, we measured a relative difference of 0.35\,\%, which we adopt as the systematic uncertainty for the PIONIER wavelength calibration. We note that this is in very good agreement with the 0.4\,\% we previously determined in \citet{Kervella_2017_01_1} from a different method.
        
        In Fig.~\ref{figure_wave} we also see a slight systematic negative offset. Although this has no impact on our previous wavelength calibration analysis, this particularly shows that the orbital solutions need to be slightly revised. As this is not the goal of the paper, we determined new orbital solutions including the new PIONIER and GRAVITY measurements in Appendix~\ref{appendix__new_orbit}.
        
        \subsection{Checking for binarity}
        \label{section__checking_for_binarity}
        
        Although our selected targets are not identified as binary stars, detecting unknown orbiting high-contrast companions are still possible when using high-angular resolution techniques. We first had a visual analysis of the data to see any variations in the $V^2$ and $CP$ measurements and cross-checked this with possible variations in the signal of the calibrators we used. A binary calibrator would lead to a bias estimate of the observables of the science target. We flagged all suspicious calibrators and reran the calibration process.
        
        We then used the \texttt{CANDID} tool \footnote{Available at \url{https://github.com/amerand/CANDID}} \citep{Gallenne_2015_07_0} on all stars of our sample to detect possible components which might bias the angular diameter determinations or/and the photometry. This is particularly important for the calibration of the SBC relation as any binary would result as an outlier. In Table~\ref{table__journal} we reported the stars for which a companion might have been detected with more than a $3\sigma$ level. We show an example is shown in Fig.~\ref{figure_chi2map} for HD45415, for which a companion is detected with a flux ratio $f \sim 1.4$\,\% in $H$. However, the possible companions are not strongly constrained with our observations, as we only have one or two brackets per epoch and a small $(u,v)$ coverage. Undetected astrometric faint companions which can bias the visible or IR photometry are still possible.
        
        \subsection{Spectroscopic observations}
        \label{section_spectroscopic_observations}
        
        We collected high-resolution echelle spectra from HARPS spectrograph located in La Silla Observatory \citep{Mayor_2003_12_0} and CHIRON spectrograph located in Cerro Tololo Observatory \citep{Tokovinin_2013_11_0}. HARPS was used in EGGS mode offering a spectral resolution of $R \sim 80000$ and CHIRON was also used with a resolution of $R \sim 80000$. Both instruments cover the spectral range $3900-6900\,\AA$. Calibrated spectra were obtained using the dedicated provided pipelines.
        
        From the reduced spectra we performed two analyses to determined the atmospheric stellar parameters. First, we determined the effective temperature $T_\mathrm{eff}$, the surface gravity $\log g$, the metallicity $[\mathrm{Fe/H}]$ and the microturbulent velocity $v_\mathrm{t}$ as described in  \citet{Villanova_2010_10_0}, in other words, using the local thermodynamic equilibrium programme MOOG \citep{Sneden_1973_09_0} and the equivalent widths (EQW) of the \ion{Fe}{I} and \ion{Fe}{II} spectral lines. As a first step, atmospheric models were calculated using ATLAS9 models \citep{Kurucz_1970__0} and initial estimates from the literature. Then, $T_\mathrm{eff}$,  $\log g$, and $v_\mathrm{t}$ were adjusted and new atmospheric models were calculated in an interactive way, in order to remove trends in excitation potential and EQWs versus abundance for $T_\mathrm{eff}$, and $v_\mathrm{t}$, respectively, and to satisfy the ionization equilibrium for $\log g$. The [Fe/H] value of the model was changed at each iteration, according to the output of the abundance analysis. A second determination of the effective temperature was also derived using the formalism of \citet{Kovtyukh_2000_06_0} based on spectral lines depth ratios and a calibration for giant stars \citep{Kovtyukh_2006_09_0}. A third estimate of the temperature was determined using $T_\mathrm{eff}-(V-K)$ calibrations \citep{Houdashelt_2000_03_0,Ramirez_2005_06_0,Worthey_2011_03_0}. Finally, we retrieved additional measurement of temperatures, gravities, metallicities and velocities from the literature, when available. We used averages and standard deviations as our final values and uncertainties.
        
        \begin{figure*}[!h]
                \centering
                \resizebox{\hsize}{!}{
                        \includegraphics[width = \linewidth]{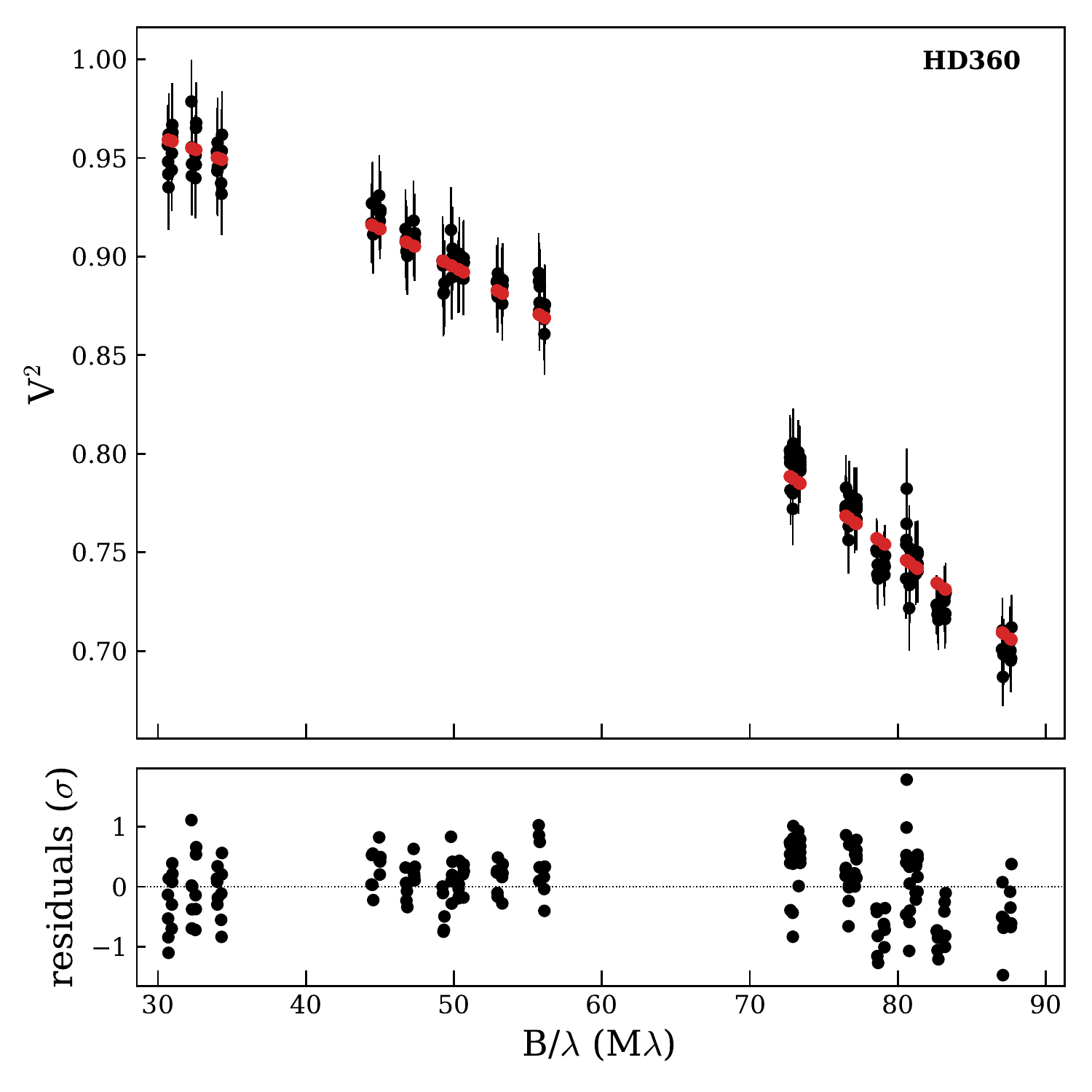}
                        \includegraphics[width = \linewidth]{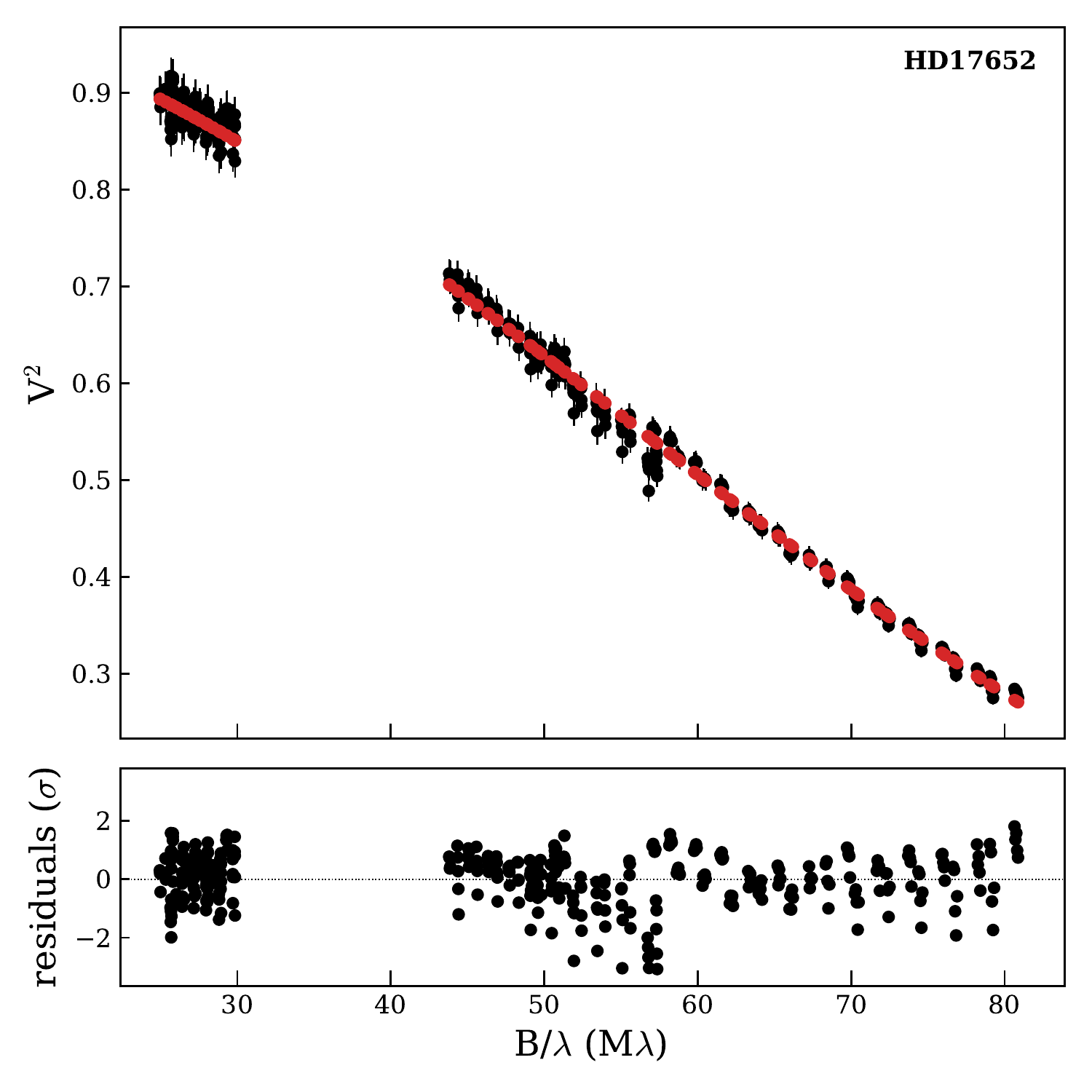}}
                %               \resizebox{\hsize}{!}{
                %                       \includegraphics[width = \linewidth]{HD15779.pdf}
                %                       \includegraphics[width = \linewidth]{HD17652.pdf}}
                \caption{Example of calibrated squared visibilities and fitted limb-darkened angular diameter model. Our measurements are represented with black dots, and the best fitted model in red.}
                \label{figure_diameters}
        \end{figure*}
        
        \section{Limb-darkened angular diameters}
        \label{section_limb_darkened_angular_diameters}
        
        We determine the limb-darkened angular diameters, $\theta_\mathrm{LD}$, for each star by fitting the calibrated squared visibilities. Assuming a circular symmetry, we followed the formalism of \citet{Merand_2015_12_0} which consist in extracting the radial intensity profile $I(r)$ of the spherical \texttt{SATLAS} models \citep{Neilson_2013_06_0}, which was converted to a visibility profile using a Hankel transform:
        \begin{equation}
        V_\lambda (x) = \frac{\int_0^1 I_\lambda(r) J_0(rx) r dr}{ \int_0^1 I_\lambda(r) r dr },
        \end{equation}
        where $\lambda$ is the wavelength, $x = \pi \theta_\mathrm{LD} B/\lambda$, $B$ is the interferometric baseline projected on to the sky, $J_0$ the Bessel function of the first kind, and $r = \sqrt{1-\mu^2}$, with $\mu = \cos(\theta)$, $\theta$ being the angle between the line of sight and a surface element of the star.
        
        The \texttt{SATLAS} grid models span effective temperatures from 3000 to 8000\,K in steps of 100\,K, effective gravities from -1 to 3 in steps of 0.25, and masses from 0.5 to 20\,$M_\odot$. For all stars, we chose the model with the closest temperature and gravity, and for a stellar mass of $1\,M_\odot$ (typical for such stars). Effective temperatures and gravities were determined as explained in Sect.~\ref{section_spectroscopic_observations}. 
        
        In Fig.~\ref{figure_diameters} we plotted the visibility curve for two stars, and all the measured angular diameters are listed in Table~\ref{table_diameters}. Errors were determined using the bootstrapping technique (with replacement) on all baselines. The listed diameters correspond to the median of the distribution and the maximum value between the 16th and 84th percentile as uncertainty. Changing the models with $T_\mathrm{eff} \pm 200$\,K, $\log g\pm 0.5$ and $M \pm 2.5\,M_\odot$ change the diameters by at most 0.3\,\%, which we added as error for each value to be conservative. Finally, we also added 0.35\,\% due to the systematic uncertainty from the wavelength calibration of PIONIER. The final overall angular diameter accuracy is better than 2.7\,\%, with a median value of 1.2\,\%.

        \begin{table*}[!h]
                \centering
                \caption{Measured and derived intrinsic stellar parameters.} 
                \begin{tabular}{cccccccc}
                        \hline\hline
                        HD  & $\theta_\mathrm{LD}$ & $\log T_\mathrm{eff}$  & $\log g$ & [Fe/H]  & $E(B-V)$  & $\log R/R_\odot$                            & $\log L/L_\odot$   \\
                        &       (mas)                                              &     ($K$)                                                   &                    & (dex)             &                       &                                                                                         &    \\
                        \hline
                        360  &  $0.906\pm0.015$  &  $3.678\pm0.002$  &  $2.62\pm0.11$  &  $-0.12\pm0.05$  &  $0.009$  &  $1.036\pm0.009$  &  $1.736\pm0.019$  \\ 
                        3750  &  $1.003\pm0.020$  &  $3.660\pm0.006$  &  $2.28\pm0.32$  &  $-0.04\pm0.07$  &  $0.002$  &  $1.022\pm0.010$  &  $1.636\pm0.019$  \\ 
                        4211  &  $1.100\pm0.011$  &  $3.656\pm0.007$  &  $2.29\pm0.23$  &  $-0.07\pm0.06$  &  $0.004$  &  $1.089\pm0.019$  &  $1.757\pm0.037$  \\ 
                        5722  &  $0.995\pm0.019$  &  $3.689\pm0.002$  &  $2.60\pm0.06$  &  $-0.17\pm0.03$  &  $0.010$  &  $1.038\pm0.011$  &  $1.785\pm0.021$  \\ 
                        8651  &  $1.228\pm0.013$  &  $3.674\pm0.003$  &  $2.50\pm0.16$  &  $-0.23\pm0.03$  &  $0.002$  &  $1.013\pm0.007$  &  $1.673\pm0.013$  \\ 
                        9362  &  $2.301\pm0.021$  &  $3.680\pm0.002$  &  $2.61\pm0.10$  &  $-0.28\pm0.10$  &  $0.000$  &  $1.033\pm0.005$  &  $1.737\pm0.011$  \\ 
                        10142  &  $0.964\pm0.006$  &  $3.674\pm0.003$  &  $2.44\pm0.04$  &  $-0.15\pm0.01$  &  $0.007$  &  $1.024\pm0.006$  &  $1.699\pm0.013$  \\ 
                        11977  &  $1.528\pm0.013$  &  $3.693\pm0.004$  &  $2.71\pm0.27$  &  $-0.24\pm0.05$  &  $0.002$  &  $1.048\pm0.006$  &  $1.819\pm0.013$  \\ 
                        12438  &  $1.091\pm0.016$  &  $3.696\pm0.006$  &  $2.43\pm0.32$  &  $-0.66\pm0.09$  &  $0.004$  &  $1.025\pm0.013$  &  $1.787\pm0.026$  \\ 
                        13468  &  $0.886\pm0.010$  &  $3.688\pm0.002$  &  $2.65\pm0.04$  &  $-0.13\pm0.05$  &  $0.009$  &  $1.048\pm0.008$  &  $1.800\pm0.015$  \\ 
                        15220  &  $1.185\pm0.016$  &  $3.651\pm0.010$  &  $2.19\pm0.04$  &  $0.26\pm0.05$  &  $0.007$  &  $1.030\pm0.016$  &  $1.615\pm0.033$  \\ 
                        15248  &  $0.949\pm0.019$  &  $3.669\pm0.003$  &  $2.45\pm0.04$  &  $0.06\pm0.05$  &  $0.010$  &  $1.041\pm0.010$  &  $1.710\pm0.020$  \\ 
                        15779  &  $1.185\pm0.014$  &  $3.684\pm0.002$  &  $2.67\pm0.04$  &  $0.00\pm0.06$  &  $0.006$  &  $1.024\pm0.007$  &  $1.737\pm0.015$  \\ 
                        16815  &  $2.248\pm0.014$  &  $3.672\pm0.006$  &  $2.65\pm0.10$  &  $-0.34\pm0.02$  &  $0.000$  &  $1.048\pm0.005$  &  $1.738\pm0.009$  \\ 
                        17652  &  $1.835\pm0.014$  &  $3.680\pm0.001$  &  $2.67\pm0.10$  &  $-0.34\pm0.10$  &  $0.001$  &  $1.019\pm0.007$  &  $1.710\pm0.014$  \\ 
                        17824  &  $1.391\pm0.015$  &  $3.700\pm0.002$  &  $2.95\pm0.03$  &  $0.08\pm0.06$  &  $0.002$  &  $0.947\pm0.009$  &  $1.647\pm0.018$  \\ 
                        18784  &  $1.036\pm0.014$  &  $3.673\pm0.003$  &  $2.35\pm0.08$  &  $-0.12\pm0.07$  &  $0.014$  &  $1.026\pm0.008$  &  $1.695\pm0.016$  \\ 
                        23319  &  $2.033\pm0.014$  &  $3.662\pm0.003$  &  $2.56\pm0.05$  &  $0.03\pm0.06$  &  $0.001$  &  $1.056\pm0.006$  &  $1.712\pm0.013$  \\ 
                        23526  &  $0.915\pm0.021$  &  $3.687\pm0.004$  &  $2.68\pm0.13$  &  $-0.15\pm0.04$  &  $0.017$  &  $1.046\pm0.012$  &  $1.792\pm0.024$  \\ 
                        23940  &  $1.093\pm0.021$  &  $3.682\pm0.004$  &  $2.43\pm0.18$  &  $-0.42\pm0.08$  &  $0.002$  &  $0.986\pm0.014$  &  $1.652\pm0.027$  \\ 
                        26464  &  $1.089\pm0.012$  &  $3.682\pm0.006$  &  $2.85\pm0.10$  &  $0.11\pm0.14$  &  $0.008$  &  $1.067\pm0.008$  &  $1.813\pm0.015$  \\ 
                        30814  &  $1.310\pm0.010$  &  $3.689\pm0.006$  &  $2.82\pm0.24$  &  $0.04\pm0.07$  &  $0.006$  &  $1.015\pm0.007$  &  $1.739\pm0.014$  \\ 
                        35369  &  $2.012\pm0.016$  &  $3.692\pm0.004$  &  $2.76\pm0.21$  &  $-0.18\pm0.02$  &  $0.000$  &  $1.062\pm0.008$  &  $1.845\pm0.017$  \\ 
                        36874  &  $1.118\pm0.011$  &  $3.664\pm0.004$  &  $2.47\pm0.07$  &  $-0.04\pm0.04$  &  $0.002$  &  $1.029\pm0.006$  &  $1.669\pm0.013$  \\ 
                        39523  &  $1.939\pm0.016$  &  $3.669\pm0.008$  &  $2.56\pm0.22$  &  $0.15\pm0.20$  &  $0.001$  &  $1.050\pm0.007$  &  $1.728\pm0.014$  \\ 
                        39640  &  $1.251\pm0.017$  &  $3.689\pm0.003$  &  $2.70\pm0.08$  &  $-0.11\pm0.03$  &  $0.006$  &  $1.031\pm0.008$  &  $1.769\pm0.016$  \\ 
                        39910  &  $1.090\pm0.008$  &  $3.659\pm0.008$  &  $2.39\pm0.21$  &  $0.18\pm0.09$  &  $0.015$  &  $1.051\pm0.006$  &  $1.693\pm0.013$  \\ 
                        40020  &  $1.012\pm0.023$  &  $3.669\pm0.008$  &  $2.43\pm0.24$  &  $0.09\pm0.08$  &  $0.013$  &  $1.050\pm0.011$  &  $1.728\pm0.023$  \\ 
                        43899  &  $1.264\pm0.017$  &  $3.658\pm0.007$  &  $2.04\pm0.24$  &  $-0.12\pm0.08$  &  $0.010$  &  $1.067\pm0.007$  &  $1.718\pm0.015$  \\ 
                        45415  &  $1.080\pm0.061$  &  $3.679\pm0.003$  &  $2.75\pm0.08$  &  $-0.02\pm0.05$  &  $0.015$  &  $1.022\pm0.026$  &  $1.715\pm0.051$  \\ 
                        46116  &  $1.145\pm0.031$  &  $3.685\pm0.003$  &  $2.48\pm0.15$  &  $-0.38\pm0.05$  &  $0.009$  &  $1.008\pm0.013$  &  $1.712\pm0.027$  \\ 
                        53629  &  $1.065\pm0.024$  &  $3.647\pm0.009$  &  $2.14\pm0.15$  &  $0.13\pm0.05$  &  $0.017$  &  $1.073\pm0.011$  &  $1.687\pm0.022$  \\ 
                        54131  &  $1.061\pm0.021$  &  $3.679\pm0.005$  &  $2.72\pm0.10$  &  $-0.10\pm0.09$  &  $0.012$  &  $1.006\pm0.010$  &  $1.684\pm0.020$  \\ 
                        56160  &  $1.411\pm0.012$  &  $3.646\pm0.008$  &  $2.19\pm0.10$  &  $0.16\pm0.09$  &  $0.010$  &  $1.092\pm0.006$  &  $1.720\pm0.012$  \\ 
                        60060  &  $0.948\pm0.010$  &  $3.683\pm0.004$  &  $2.58\pm0.14$  &  $-0.11\pm0.03$  &  $0.018$  &  $1.038\pm0.007$  &  $1.761\pm0.014$  \\ 
                        60341  &  $1.190\pm0.022$  &  $3.665\pm0.006$  &  $2.42\pm0.26$  &  $0.06\pm0.07$  &  $0.010$  &  $1.062\pm0.010$  &  $1.737\pm0.019$  \\ 
                        62412  &  $0.950\pm0.014$  &  $3.692\pm0.003$  &  $2.76\pm0.09$  &  $0.03\pm0.03$  &  $0.013$  &  $1.014\pm0.008$  &  $1.751\pm0.017$  \\ 
                        62713  &  $1.446\pm0.012$  &  $3.666\pm0.004$  &  $2.42\pm0.33$  &  $0.09\pm0.05$  &  $0.005$  &  $1.015\pm0.006$  &  $1.645\pm0.013$  \\ 
                        68312  &  $1.020\pm0.023$  &  $3.704\pm0.002$  &  $2.75\pm0.05$  &  $-0.10\pm0.01$  &  $0.011$  &  $0.974\pm0.012$  &  $1.718\pm0.023$  \\ 
                        74622  &  $1.020\pm0.015$  &  $3.647\pm0.005$  &  $2.26\pm0.20$  &  $-0.03\pm0.03$  &  $0.013$  &  $1.026\pm0.008$  &  $1.593\pm0.015$  \\ 
                        75916  &  $1.013\pm0.021$  &  $3.671\pm0.003$  &  $2.47\pm0.11$  &  $0.15\pm0.05$  &  $0.008$  &  $1.064\pm0.010$  &  $1.764\pm0.021$  \\ 
                        176704  &  $1.317\pm0.012$  &  $3.655\pm0.004$  &  $2.56\pm0.10$  &  $0.36\pm0.10$  &  $0.007$  &  $1.049\pm0.011$  &  $1.671\pm0.021$  \\ 
                        177873  &  $1.958\pm0.029$  &  $3.667\pm0.003$  &  $2.59\pm0.10$  &  $0.01\pm0.10$  &  $0.002$  &  $1.051\pm0.010$  &  $1.722\pm0.020$  \\ 
                        188887  &  $1.595\pm0.011$  &  $3.650\pm0.003$  &  $2.45\pm0.10$  &  $0.11\pm0.10$  &  $0.005$  &  $1.093\pm0.005$  &  $1.739\pm0.010$  \\ 
                        191584  &  $1.024\pm0.022$  &  $3.649\pm0.008$  &  $2.35\pm0.15$  &  $0.22\pm0.00$  &  $0.009$  &  $1.047\pm0.011$  &  $1.645\pm0.021$  \\ 
                        204381  &  $1.524\pm0.017$  &  $3.703\pm0.001$  &  $2.96\pm0.10$  &  $-0.01\pm0.10$  &  $0.001$  &  $0.934\pm0.008$  &  $1.635\pm0.016$  \\ 
                        219784  &  $2.117\pm0.025$  &  $3.661\pm0.003$  &  $2.29\pm0.17$  &  $-0.10\pm0.04$  &  $0.001$  &  $1.065\pm0.009$  &  $1.725\pm0.019$  \\ 
                        220572  &  $1.092\pm0.013$  &  $3.674\pm0.002$  &  $2.64\pm0.09$  &  $0.07\pm0.01$  &  $0.003$  &  $1.028\pm0.007$  &  $1.705\pm0.014$  \\ 
                        \hline
                \end{tabular}
                \tablefoot{
                        Data from the literature were also used to the average estimates of $T_\mathrm{eff}$, $\log g$ and [Fe/H] \citep{Liu_2007_12_0,Jofre_2015_02_0,Alves_2015_04_0,Jones_2011_12_0,Mishenina_2006_09_0,Mikolaitis_2017_04_0,Feuillet_2016_01_0,Allende-Prieto_1999_12_0,Proust_1988_06_0,Mishenina_2006_09_0}. $E(B-V)$ were determined as explained in \citet{Suchomska_2015_07_0}.}
                \label{table_diameters}
        \end{table*}
        
        \section{Stellar properties}
        \label{section__stellar_properties}
        
        \subsection{Stellar radii and luminosities}
        
        From the measured angular diameters and Gaia parallaxes (or from Hipparcos if not in the Gaia DR2, see Table~\ref{table__journal}), we can derive the stellar radii and the luminosities through the following equations
        
        \begin{equation}
        R[R_\odot] = 107.523 \frac{\theta_\mathrm{LD}[\mathrm{mas}]}{\pi[\mathrm{mas}]},
        \end{equation}
        \begin{equation}
        \frac{L}{L_\odot} = \left(\frac{R}{R_\odot}\right)^2 \left(\frac{T_\mathrm{eff}}{T_\mathrm{eff,\odot}}\right)^4
        \end{equation}
        
        The values are listed in Table~\ref{table_diameters}. For the conversions, we adopted the nominal solar and astronomical constants from IAU 2015 Resolution B3 \citep{Prsa_2016_08_0} and CODATA values \citep{Mohr_2016_07_0}. Gaia parallaxes were corrected from the zero point offset of $\sim0.03$\,mas,  and we quadratically added to the uncertainties a (conservative) systematic error of $\pm 0.1$\,mas \citep{Gaia-Collaboration_2018_04_0}.
        
        \subsection{Ages and masses}
        
        We used the \texttt{PARSEC} \citep{Bressan_2012_11_0} and \texttt{BaSTI} \citep{Pietrinferni_2004_09_0} isochrone models to estimate the stellar masses and ages. These models are well suited as they include the horizontal and asymptotic giant branch evolutionary phases, and contain a wide range of initial masses and metallicities. In addition, it enable us to test the uncertainty of age and mass estimate induced by different stellar models.
        
        \texttt{PARSEC} models are computed for a scaled-solar composition with $Z_\odot=0.0152$, follow a helium initial content relation $Y_\mathrm{i} = 0.2485 + 1.78\,Z_\mathrm{i}$, and include moderate convective core overshooting. The \texttt{BaSTI} models are computed for a scaled-solar composition with $Z_\odot=0.0198$, a model composition following $\Delta Y/\Delta Z \sim 1.4$ with $Y = 0.245$ at $Z = 0$, and also include convective core overshooting. Both models assume the Reimers mass-loss rate $\eta = 0.2$.
        
        For our fitting procedure, we computed several isochrones from the \texttt{PARSEC} database tool\footnote{\url{http://stev.oapd.inaf.it/cgi-bin/cmd}}, with ages ranging from $t = 0.1$ to 13\,Gyr by step of 0.01\,Myr, and metallicities from $Z = 0.003$ to 0.06 (i.e. $-0.7 < \mathrm{[Fe/H]} < +0.6$, using $[\mathrm{Fe/H}] \sim \log{(Z/Z_\odot)}$), by step of 0.001. The \texttt{BaSTI} isochrones are pre-computed in their database\footnote{\url{http://basti.oa-teramo.inaf.it/index.html}}, we downloaded models for $t = 0.1-10$\,Myr by step of 0.01\,Myr and $Z = 0.002, 0.004, 0.008, 0.01, 0.0198, 0.03$ and 0.04 (i.e. $-1.0 < \mathrm{Fe/H} < 0.3$). These models are also for a scaled-solar composition (with $Z_\odot=0.0198$) and also include overshooting. HD176704 has $\mathrm{Fe/H} = 0.36$\,dex, above the range of the \texttt{BaSTI} isochrones, but we rounded it down to 0.3. We chose grids fine enough in age to avoid interpolation (which might cause problems); the closest age is therefore always chosen in our fitting procedure.
        
        The \texttt{PARSEC} output tables provide the luminosities, effective temperatures, effective gravities, and masses. We computed the linear radius from the table values following the equation
        \begin{equation}
        \log \left( \dfrac{R}{R_\odot} \right) = \dfrac{1}{2} \log \left( \dfrac{L}{L_\odot} \right) - 2\log T_\mathrm{eff} + 2\log T_\odot.
        \end{equation}
        The surface gravities for \texttt{BaSTI} were determined using Newton's law of universal gravitation
        \begin{equation}
        \log g = \log \left( \dfrac{M}{M_\odot} \right) + 4\log T_\mathrm{eff} - 4\log T_\odot - \log \left( \dfrac{L}{L_\odot} \right) + \log g_\odot,
        \end{equation}
        with the solar constants from \citet{Prsa_2016_08_0}.
        
        Then, from these grids, we performed our isochrone fits by adopting fixed values of metallicity, and searched for the best age fit in luminosity, effective temperature, radii and effective gravity following a $\chi^2$ statistic, that is, minimizing
        \begin{equation}
        \chi^2 = \sum \left[ \left(\dfrac{\Delta L}{\sigma_\mathrm{L}}\right)^2 + \left(\dfrac{\Delta T}{\sigma_\mathrm{T}}\right)^2 + 
        \left(\dfrac{\Delta \log g}{\sigma_\mathrm{\log g}}\right)^2 + \left(\dfrac{\Delta R}{\sigma_\mathrm{R}}\right)^2 \right]
        \end{equation}
        
        Our fitting procedure was the following. For the \texttt{PARSEC} isochrones, we first chose the closest grid in $Z$ for a given metallicity (given in Table~\ref{table_diameters}). We note that the grid was not interpolated as our downloaded tracks are also fine enough in metallicity. Then, we searched for the global $\chi^2$ minimum in age and mass by fitting all isochrones for that given metallicity. A second fit is then performed around that global minimum values. For the \texttt{BaSTI} models, which are unfortunately not fine enough in metallicity, we first interpolated all isochrones to the given metallicity. We then also searched for the global $\chi^2$ minimum in age and mass by fitting all isochrones for that given metallicity. A second fit was also performed around that global minimum values. 
        
\begin{figure}[!h]
                \centering
                \resizebox{\hsize}{!}{\includegraphics[width = \linewidth]{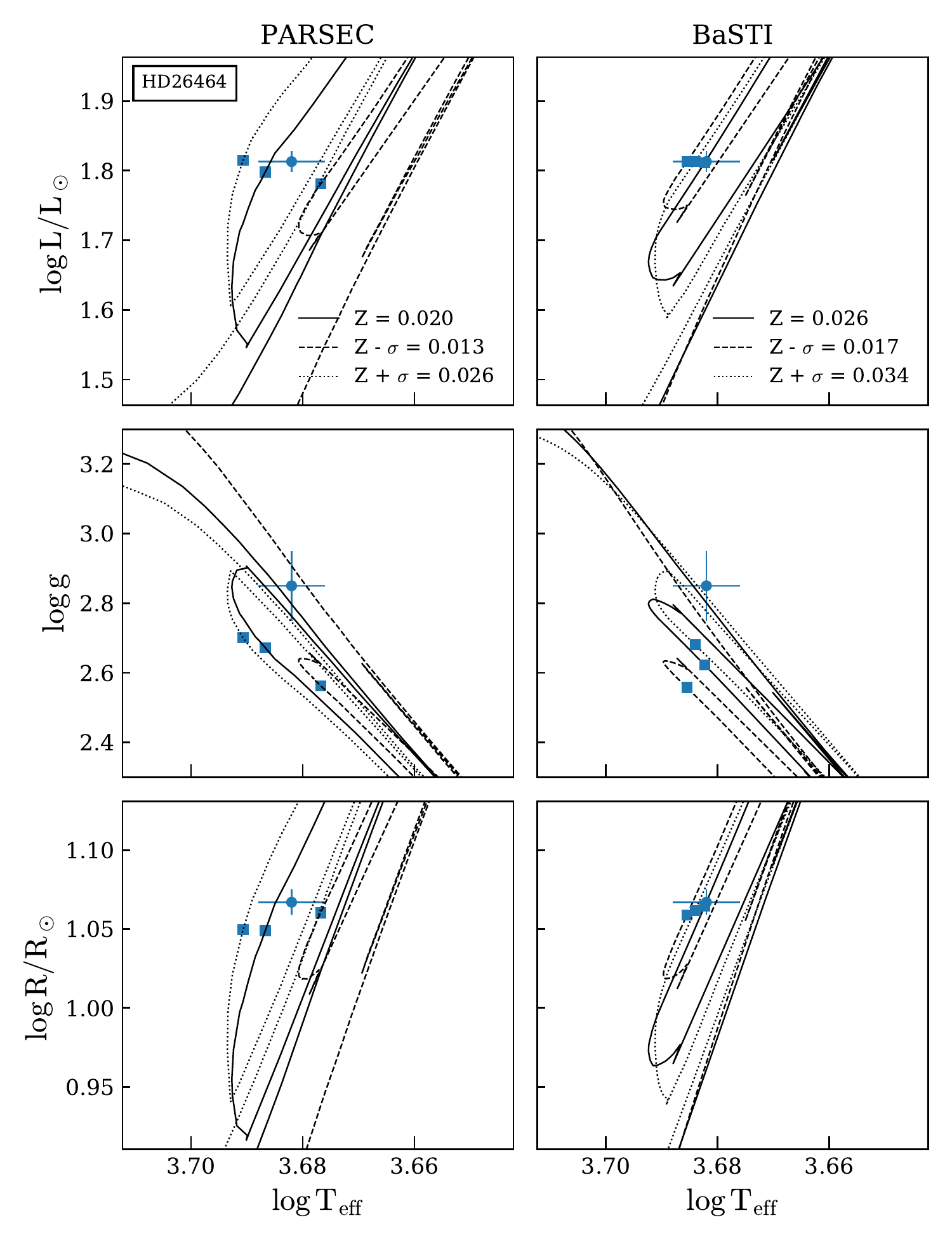}}
                \caption{Fitted \texttt{PARSEC} and \texttt{BaSTI} isochrones for HD26464.}
                \label{figure_isochrones}
        \end{figure}

        To assess the uncertainties for the \texttt{PARSEC} and \texttt{BaSTI} models, we repeated the process with $Z \pm \sigma$. The final age and mass corresponding to each isochrone model are listed in Table~\ref{table__age}. We also listed the average and standard deviation between both models, together with the corresponding evolutionary status of the star. Figure~\ref{figure_isochrones} shows an example of a fitted isochrones for the star HD26464.

        \begin{table*}[!h]
                \centering
                \caption{Estimated stellar mass, age and evolutionary status of our giant stars.}
                \begin{tabular}{c|cc|cc|cc}
                        \hline\hline
                        &       \multicolumn{2}{c|}{\texttt{PARSEC}}  &       \multicolumn{2}{c|}{\texttt{BaSTI}}  &  \multicolumn{2}{c}{Average}  \\
                        Star                    & Age  & Mass                                           & Age  & Mass     & Age  & Mass                                    \\
                        & (Gyr) &     ($M/M\odot$)     & (Gyr) &     ($M/M\odot$)  & (Gyr) &     ($M/M\odot$)    \\
                        \hline
                        HD360  &  $1.71 \pm 0.10$ &  $1.72 \pm 0.05$    &       $1.75 \pm 0.20$   &  $1.63 \pm 0.06$  &       $1.73 \pm 0.02$ &  $1.68 \pm 0.04$      \\ %& RGB  \\
                        HD3750 &  $6.31 \pm 1.19$  &  $1.16 \pm 0.07$   &       $7.33 \pm 1.65$ &  $1.11 \pm 0.10$    &       $6.82 \pm 0.51$ &  $1.14 \pm 0.03$ \\ %& RGB       \\
                        HD4211 &  $4.64 \pm 1.01$  &  $1.26 \pm 0.09$   &       $6.50 \pm 1.22$   &  $1.12 \pm 0.06$ &        $5.57 \pm 0.93$ &  $1.19 \pm 0.07$      \\ %& RGB  \\      
                        HD5722 &  $2.24 \pm 0.05$  &  $1.57 \pm 0.01$   &       $2.17 \pm 0.12$  &  $1.50 \pm 0.04$   &       $2.20 \pm 0.04$ &  $1.53 \pm 0.04$      \\ %& RC   \\              
                        HD8651 &  $3.55 \pm 0.33$  &  $1.32 \pm 0.05$   &       $4.08 \pm 0.31$  &  $1.26 \pm 0.03$   &       $3.82 \pm 0.27$ &  $1.29 \pm 0.03$      \\ %& RGB  \\      
                        HD9362 &  $2.61 \pm 0.75$  &  $1.45 \pm 0.16$   &       $2.33 \pm 0.51$  &  $1.47 \pm 0.14$   &       $2.47 \pm 0.14$ &  $1.46 \pm 0.01$      \\ %& RGB  \\                      
                        HD10142 &  $7.08 \pm 0.05$ &  $1.04 \pm 0.10$   &       $3.50 \pm 0.10$  &  $1.32 \pm 0.01$   &       $5.29 \pm 1.79$ &  $1.18 \pm 0.14$      \\ %& RC/RGB       \\              
                        HD11977 &  $1.78 \pm 0.89$  &  $1.70 \pm 0.31$  &       $0.87 \pm 0.05$  &  $2.08 \pm 0.04$   &       $1.32 \pm 0.46$ &  $1.89 \pm 0.19$      \\ %& RC   \\              
                        HD12438 &  $2.61 \pm 0.71$  &  $1.36 \pm 0.14$  &       $3.33 \pm 1.90$  &  $1.28 \pm 0.22$   &       $2.97 \pm 0.36$ &  $1.32 \pm 0.04$      \\ %& RGB  \\              
                        HD13468 &  $1.04 \pm 0.06$  &  $2.04 \pm 0.06$  &       $0.93 \pm 0.05$  &  $2.03 \pm 0.04$   &       $0.99 \pm 0.05$ &  $2.04 \pm 0.01$      \\ %& RC/RGB       \\      
                        HD15220 &  $12.60 \pm 0.05$  &  $0.92 \pm 0.01$ &       $9.50 \pm 0.10$  &  $1.10 \pm 0.01$   &       $11.05 \pm 1.55$ &  $1.01 \pm 0.09$ \\ %& RC/RGB    \\                      
                        HD15248   &  $4.64 \pm 0.25$  &  $1.26 \pm 0.01$&       $5.33 \pm 0.62$  &  $1.18 \pm 0.05$   &       $4.99 \pm 0.35$ &  $1.22 \pm 0.04$      \\ %& RC   \\              
                        HD15779 &  $1.41 \pm 0.35$  &  $1.90 \pm 0.13$  &       $1.42 \pm 0.24$  &  $1.80 \pm 0.03$   &       $1.41 \pm 0.01$ &  $1.85 \pm 0.05$      \\ %& RC/RGB       \\              
                        HD16815 &  $3.55 \pm 0.05$  &  $1.29 \pm 0.01$  &       $4.33 \pm 0.24$  &  $1.22 \pm 0.02$   &       $3.94 \pm 0.39$ &  $1.25 \pm 0.04$ \\ %& RGB       \\              
                        HD17652 &  $3.29 \pm 1.08$  &  $1.34 \pm 0.17$  &       $3.25 \pm 0.94$  &  $1.33 \pm 0.14$   &       $3.27 \pm 0.02$ &  $1.33 \pm 0.01$      \\ %& RGB  \\              
                        HD17824 &  $0.76 \pm 0.04$  &  $2.36 \pm 0.06$  &       $0.67 \pm 0.05$  &  $2.42 \pm 0.06$   &       $0.72 \pm 0.05$ &  $2.39 \pm 0.03$      \\ %& RGB  \\              
                        HD18784 &  $10.80 \pm 1.18$  &  $0.90 \pm 0.03$ &       $7.58 \pm 2.71$  &  $1.04 \pm 0.18$   &       $9.19 \pm 1.61$ &  $0.97 \pm 0.07$      \\ %& RC   \\              
                        HD23319 &  $2.82 \pm 0.46$  &  $1.51 \pm 0.09$  &       $3.17 \pm 0.42$  &  $1.42 \pm 0.09$   &       $2.99 \pm 0.17$ &  $1.46 \pm 0.05$      \\ %& RGB  \\              
                        HD23526 &  $1.04 \pm 0.06$  &  $2.03 \pm 0.04$  &       $0.97 \pm 0.05$  &  $2.00 \pm 0.04$  & $1.00 \pm 0.04$ &  $2.02 \pm 0.01$     \\ %& RGB  \\      
                        HD23940 &  $3.98 \pm 0.99$  &  $1.24 \pm 0.11$  &       $4.67 \pm 1.03$  &  $1.16 \pm 0.10$  & $4.32 \pm 0.34$ &  $1.20 \pm 0.04$     \\ %& RGB  \\      
                        HD26464 &  $1.31 \pm 0.40$  &  $2.09 \pm 0.28$  &       $1.33 \pm 0.31$  &  $2.02 \pm 0.28$  & $1.32 \pm 0.01$ &  $2.06 \pm 0.03$     \\ %& RC   \\      
                        HD30814 &  $1.52 \pm 0.17$  &  $1.96 \pm 0.13$  &       $1.32 \pm 0.45$  &  $1.96 \pm 0.31$  & $1.42 \pm 0.10$ &  $1.96 \pm 0.01$     \\ %& RC   \\      
                        HD35369 &  $1.92 \pm 0.42$  &  $1.82 \pm 0.08$  &       $1.42 \pm 0.12$  &  $1.82 \pm 0.08$  & $1.67 \pm 0.25$ &  $1.77 \pm 0.05$     \\ %& RC   \\      
                        HD36874 &  $3.69 \pm 0.20$  &  $1.36 \pm 0.03$  &       $4.42 \pm 0.51$  &  $1.24 \pm 0.04$  & $4.05 \pm 0.36$ &  $1.30 \pm 0.06$     \\ %& RGB  \\      
                        HD39523 &  $2.93 \pm 0.69$  &  $1.52 \pm 0.16$  &       $3.25 \pm 0.61$  &  $1.41 \pm 0.12$  & $3.09 \pm 0.16$ &  $1.47 \pm 0.06$     \\ %& RC   \\      
                        HD39640 &  $1.59 \pm 0.40$  &  $1.81 \pm 0.13$  &       $0.93 \pm 0.05$  &  $2.03 \pm 0.04$  & $1.26 \pm 0.33$ &  $1.92 \pm 0.11$     \\ %& RC/RGB       \\      
                        HD39910 &  $5.62 \pm 2.80$  &  $1.23 \pm 0.21$  &       $5.92 \pm 1.53$  &  $1.19 \pm 0.13$  & $5.77 \pm 0.15$ &  $1.21 \pm 0.02$     \\ %& RC   \\      
                        HD40020 &  $6.07 \pm 1.65$  &  $1.17 \pm 0.12$  &       $6.00 \pm 2.45$  &  $1.18 \pm 0.17$  & $6.03 \pm 0.03$ &  $1.17 \pm 0.01$     \\ %& RC   \\      
                        HD43899 &  $4.64 \pm 1.01$  &  $1.25 \pm 0.10$  &       $8.00 \pm 1.47$  &  $1.06 \pm 0.06$  & $6.32 \pm 1.68$ &  $1.15 \pm 0.10$     \\ %& RGB  \\      
                        HD45415 &  $1.53 \pm 0.41$  &  $1.84 \pm 0.15$  &       $1.50 \pm 0.35$  &  $1.78 \pm 0.22$  & $1.51 \pm 0.01$ &  $1.81 \pm 0.03$     \\ %& RC   \\      
                        HD46116 &  $2.71 \pm 0.15$  &  $1.40 \pm 0.03$  &       $2.33 \pm 0.31$  &  $1.44 \pm 0.09$  & $2.52 \pm 0.19$ &  $1.42 \pm 0.02$     \\ %& RGB  \\      
                        HD53629 &  $8.91 \pm 0.05$  &  $1.09 \pm 0.01$  &       $9.17 \pm 0.47$  &  $1.08 \pm 0.01$  & $9.04 \pm 0.13$ &  $1.08 \pm 0.01$     \\ %& RGB  \\      
                        HD54131 &  $1.85 \pm 0.36$  &  $1.69 \pm 0.14$  &       $1.75 \pm 0.20$  &  $1.63 \pm 0.06$  & $1.80 \pm 0.05$ &  $1.66 \pm 0.03$     \\ %& RC   \\      
                        HD56160 &  $9.61 \pm 2.08$  &  $1.06 \pm 0.09$  &       $7.50 \pm 1.63$  &  $1.16 \pm 0.09$  & $8.56 \pm 1.06$ &  $1.10 \pm 0.06$     \\ %& RC   \\      
                        HD60060 &  $2.82 \pm 0.26$  &  $1.47 \pm 0.06$  &       $2.00 \pm 0.54$  &  $1.57 \pm 0.19$  & $2.41 \pm 0.41$ &  $1.52 \pm 0.05$     \\ %& RC   \\      
                        HD60341 &  $5.20 \pm 3.43$  &  $1.25 \pm 0.29$  &       $3.17 \pm 0.96$  &  $1.42 \pm 0.15$  & $4.19 \pm 1.02$ &  $1.33 \pm 0.09$     \\ %& RC   \\      
                        HD62412 &  $0.83 \pm 0.04$  &  $2.29 \pm 0.05$  &       $1.07 \pm 0.26$  &  $2.15 \pm 0.07$  & $0.95 \pm 0.12$ &  $2.22 \pm 0.07$     \\ %& RC   \\      
                        HD62713 &  $4.47 \pm 3.28$  &  $1.34 \pm 0.31$  &       $2.83 \pm 0.31$  &  $1.49 \pm 0.05$  & $3.65 \pm 0.82$ &  $1.41 \pm 0.08$     \\ %& RGB  \\      
                        HD68312 &  $1.41 \pm 0.05$  &  $1.96 \pm 0.01$  &       $1.25 \pm 0.05$  &  $1.91 \pm 0.01$  & $1.33 \pm 0.08$ &  $1.94 \pm 0.03$     \\ %& RC   \\      
                        HD74622 &  $12.11 \pm 0.67$  &  $0.97 \pm 0.01$ &       $9.50 \pm 0.01$  &  $1.00 \pm 0.01$  & $10.81 \pm 1.31$ &  $0.99 \pm 0.02$    \\ %& RGB  \\      
                        HD75916 &  $2.82 \pm .26$  &  $1.56 \pm 0.07$   &       $2.92 \pm 0.42$  &  $1.47 \pm 0.09$  & $2.87 \pm 0.05$ &  $1.52 \pm 0.05$     \\ %& RC   \\      
                        HD176704 &  $2.61 \pm 0.28$  &  $1.62 \pm 0.06$ &       $3.17 \pm 0.42$  &  $1.48 \pm 0.07$  & $2.89 \pm 0.28$ &  $1.55 \pm 0.07$     \\ %& RGB  \\      
                        HD177873 &  $2.93 \pm 0.97$  &  $1.48 \pm 0.18$ &       $2.67 \pm 0.62$  &  $1.51 \pm 0.14$  & $2.80 \pm 0.13$ &  $1.50 \pm 0.01$     \\ %& RC/RGB       \\      
                        HD188887   &  $3.29 \pm 0.36$  &  $1.45 \pm 0.03$       &       $5.83 \pm 1.65$  &  $1.23 \pm 0.14$  & $4.56 \pm 1.27$ &  $1.34 \pm 0.11$     \\ %& RGB  \\      
                        HD191584   &  $5.62 \pm 0.05$  &  $1.25 \pm 0.01$       &       $6.50 \pm 0.05$  &  $1.22 \pm 0.01$  & $6.06 \pm 0.44$ &  $1.23 \pm 0.01$     \\ %& RGB  \\      
                        HD204381 &  $1.00 \pm 0.25$  &  $2.15 \pm 0.17$ &       $0.77 \pm 0.17$  &  $2.26 \pm 0.19$  & $0.88 \pm 0.12$ &  $2.20 \pm 0.06$     \\ %& RGB  \\      
                        HD219784 &  $3.98 \pm 0.37$  &  $1.31 \pm 0.05$ &       $5.83 \pm 0.62$  &  $1.15 \pm 0.03$  & $4.91 \pm 0.93$ &  $1.23 \pm 0.08$     \\ %& RGB  \\      
                        HD220572 &  $1.64 \pm 0.09$  &  $1.81 \pm 0.04$ &       $1.50 \pm 0.05$  &  $1.83 \pm 0.01$  & $1.57 \pm 0.07$ &  $1.82 \pm 0.03$     \\ %& RGB  \\      
                        \hline
                \end{tabular}
                \label{table__age}
        \end{table*}
        
        \begin{figure}[!h]
                \centering
                \resizebox{\hsize}{!}{\includegraphics[width = \linewidth]{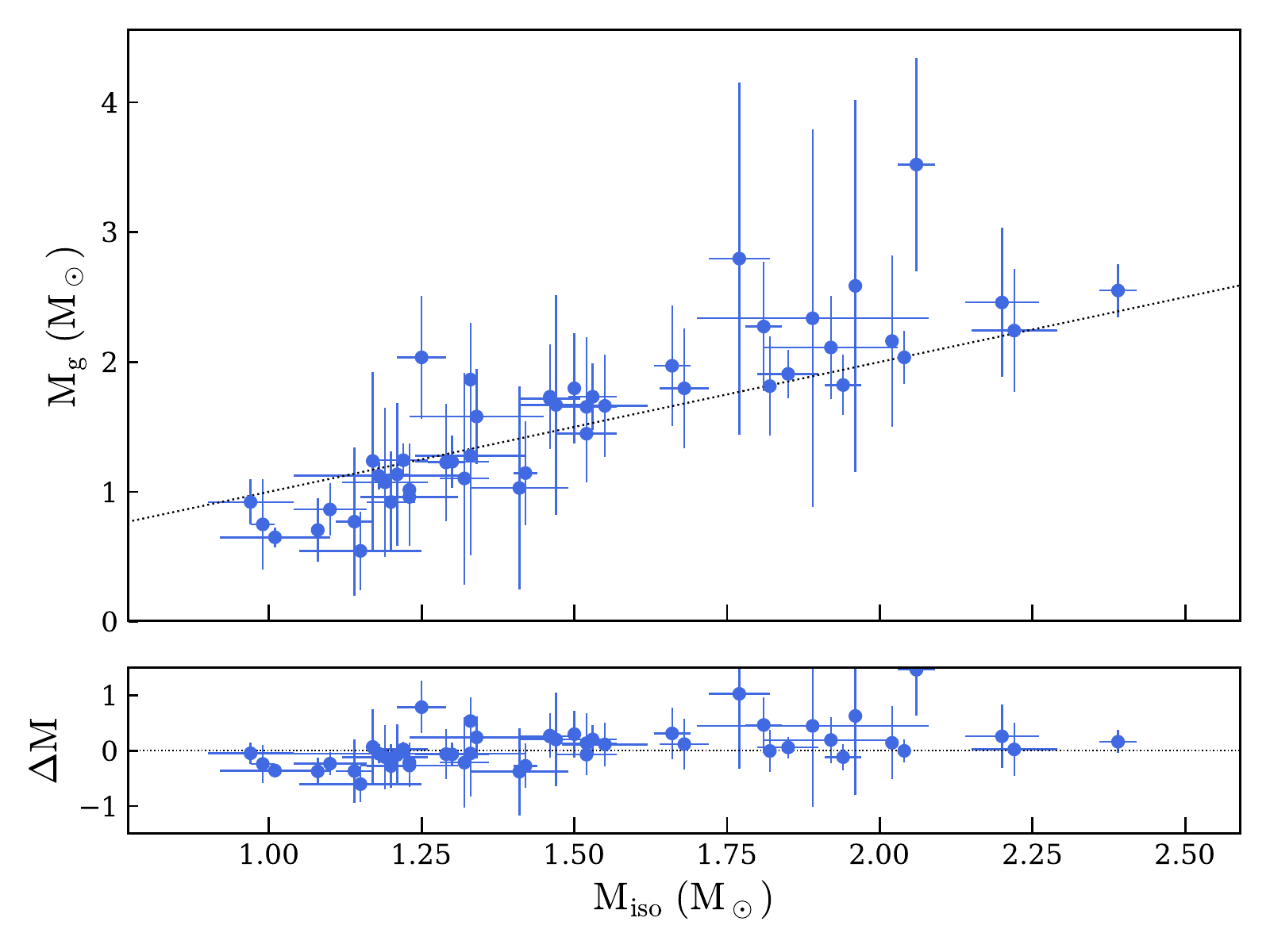}}
                \caption{Masses $M_\mathrm{iso}$ derived from isochrone fitting compared to calculated ones $M_\mathrm{g}$. The dotted line denotes a 1:1 relation.}
                \label{figure_gravity_masses}
        \end{figure}
        
        In most cases, both models give similar age and mass values, within the uncertainties. We found masses in the range $0.97 < M/M_\odot < 2.39$ and ages $0. 72 < t < 11.05$\,Gyr, which is consistent with what we expect from such stars. We can see that the masses are better determined than the ages. Comparison of the parameters between the two isochrone sets reveals no obvious systematic trends, but in some cases there are significant differences in age.
        
        The average masses determined from isochrones fitting can be compared to the ones calculated from the surface gravity, that is, from Newton's law of universal gravitation
        \begin{equation}
        \dfrac{M_\mathrm{g}}{M_\odot} = \dfrac{g}{g_\odot} \left(\dfrac{R}{R_\odot}\right)^2,
        \end{equation}
        with $g$ the surface gravity and $R$ the stellar radius. This is presented in Fig.~\ref{figure_gravity_masses}. No specific trend or offset is detected and they are in rather good agreement with each other, within 1-2$\sigma$.
        
        Our derived values are also consistent with previous works of \citet{Luck_2015_09_0} who used three different isochrone models \citep{Bertelli_1994_08_0,Demarque_2004_12_0,Dotter_2008_09_0} to estimate the mass and age of some stars in our sample. The comparison with our derived mean values are listed in Table~\ref{table__comparison_ages}. We can see that in some cases, models give very different results in age, while the mass values are less scattered. We mention here that accurate angular diameter measurements help in better constraining isochrones via multi-observables fitting, although it still depends on accurate metallicity determinations.
        
        \begin{table*}[!h]
                \centering
                \caption{Comparison with previous works with different isochrone models from \citet{Luck_2015_09_0}.}
                \begin{tabular}{c|cccc|cccc}
                        \hline\hline
                        &       \multicolumn{4}{c|}{Age (Gyr)}  &       \multicolumn{4}{c}{Mass  ($M/M\odot$)}   \\
                        Star                    & B  & D &Y & G         & B  & D &Y & G                                            \\
                        %& (Gyr) &     ($M/M\odot$)     & (Gyr) &     ($M/M\odot$)  & (Gyr) &     ($M/M\odot$)    \\
                        \hline
                        HD360  &  2.02  &  5.63  &  1.73  &  1.73  &  1.78  &  1.35  &  1.91  &  1.68 \\
                        HD4211  &  3.53  &  3.25  &  3.95  &  5.57  &  1.49  &  1.46  &  1.51  &  1.19 \\
                        HD5722  &  4.06  &  2.39  &  1.60  &  2.20  &  1.29  &  1.65  &  1.74  &  1.53 \\
                        HD9362  &  3.33  &  6.25  &  1.50  &  2.47  &  1.42  &  1.08  &  1.89  &  1.46 \\
                        HD11977  &  1.35  &  1.08  &  0.80  &  1.32  &  1.91  &  2.18  &  2.32  &  1.89 \\
                        HD12438  &  3.19  &  1.08  &  1.87  &  2.97  &  1.40  &  2.18  &  1.70  &  1.32 \\
                        HD15779  &  5.35  &  5.50  &  1.10  &  1.41  &  1.23  &  1.12  &  2.18  &  1.85 \\
                        HD16815  &  4.76  &  - &  3.47  &  3.94  &  1.34  &  - &  1.49  &  1.25 \\
                        HD17824  &  0.71  &  1.53  &  - &  0.72  &  2.30  &  1.87  &  1.93  &  2.39 \\
                        HD18784  &  2.99  &  6.18  &  1.20  &  9.19  &  1.60  &  1.21  &  2.13  &  0.97 \\
                        HD23319  &  3.70  &  1.61  &  1.90  &  2.99  &  1.48  &  1.94  &  1.83  &  1.46 \\
                        HD23526  &  5.12  &  6.00  &  1.74  &  1.00  &  1.24  &  1.09  &  1.82  &  2.02 \\
                        HD23940  &  3.65  &  8.00  &  1.85  &  4.32  &  1.57  &  1.00  &  1.77  &  1.20 \\
                        HD30814  &  4.01  &  3.07  &  1.20  &  1.42  &  1.44  &  1.58  &  2.04  &  1.96 \\
                        HD35369  &  4.20  &  2.54  &  - &  1.67  &  1.33  &  1.61  &  1.83  &  1.77 \\
                        HD39640  &  2.93  &  3.25  &  - &  1.26  &  1.51  &  1.31  &  - &  1.92 \\
                        HD39910  &  3.49  &  2.54  &  1.35  &  5.77  &  1.44  &  1.63  &  2.04  &  1.21 \\
                        HD40020  &  2.41  &  2.31  &  3.04  &  6.03  &  1.77  &  1.63  &  1.67  &  1.17 \\
                        HD45415  &  2.34  &  6.38  &  1.13  &  1.51  &  1.68  &  1.07  &  2.16  &  1.81 \\
                        HD46116  &  - &  2.35  &  1.10  &  2.52  &  1.81  &  1.71  &  2.08  &  1.42 \\
                        HD54131  &  3.67  &  3.57  &  1.63  &  1.80  &  1.44  &  1.45  &  1.85  &  1.66 \\
                        HD60341  &  2.13  &  2.88  &  4.94  &  4.19  &  1.78  &  1.53  &  1.40  &  1.33 \\
                        HD62412  &  2.06  &  2.18  &  - &  0.95  &  1.74  &  1.67  &  1.93  &  2.22 \\
                        HD62713  &  - &  2.75  &  3.60  &  3.65  &  - &  1.53  &  1.62  &  1.41 \\
                        HD68312  &  0.60  &  4.70  &  - &  1.33  &  2.45  &  1.52  &  2.22  &  1.94 \\
                        HD75916  &  4.83  &  3.88  &  4.60  &  2.87  &  1.31  &  1.39  &  1.50  &  1.52 \\
                        HD176704  &  2.62  &  3.00  &  - &  2.89  &  1.60  &  1.55  &  1.84  &  1.55 \\
                        HD204381  &  0.90  &  3.94  &  - &  0.88  &  2.06  &  1.64  &  2.13  &  2.20 \\
                        HD219784  &  2.65  &  2.19  &  2.57  &  4.91  &  1.72  &  1.64  &  1.76  &  1.23 \\
                        \hline
                \end{tabular}
                \label{table__comparison_ages}
                \tablefoot{B, D, Y, and G stands for the Bertelli isochrones \citep{Bertelli_1994_08_0}, the Dartmouth isochrones \citep{Dotter_2008_09_0}, the Y$^2$ isochrones \citep{Demarque_2004_12_0} and this work, respectively.}
        \end{table*}
        
        \subsection{Age-metallicity relation}
        
\begin{figure}[!h]
                \centering
                \resizebox{\hsize}{!}{\includegraphics[width = \linewidth]{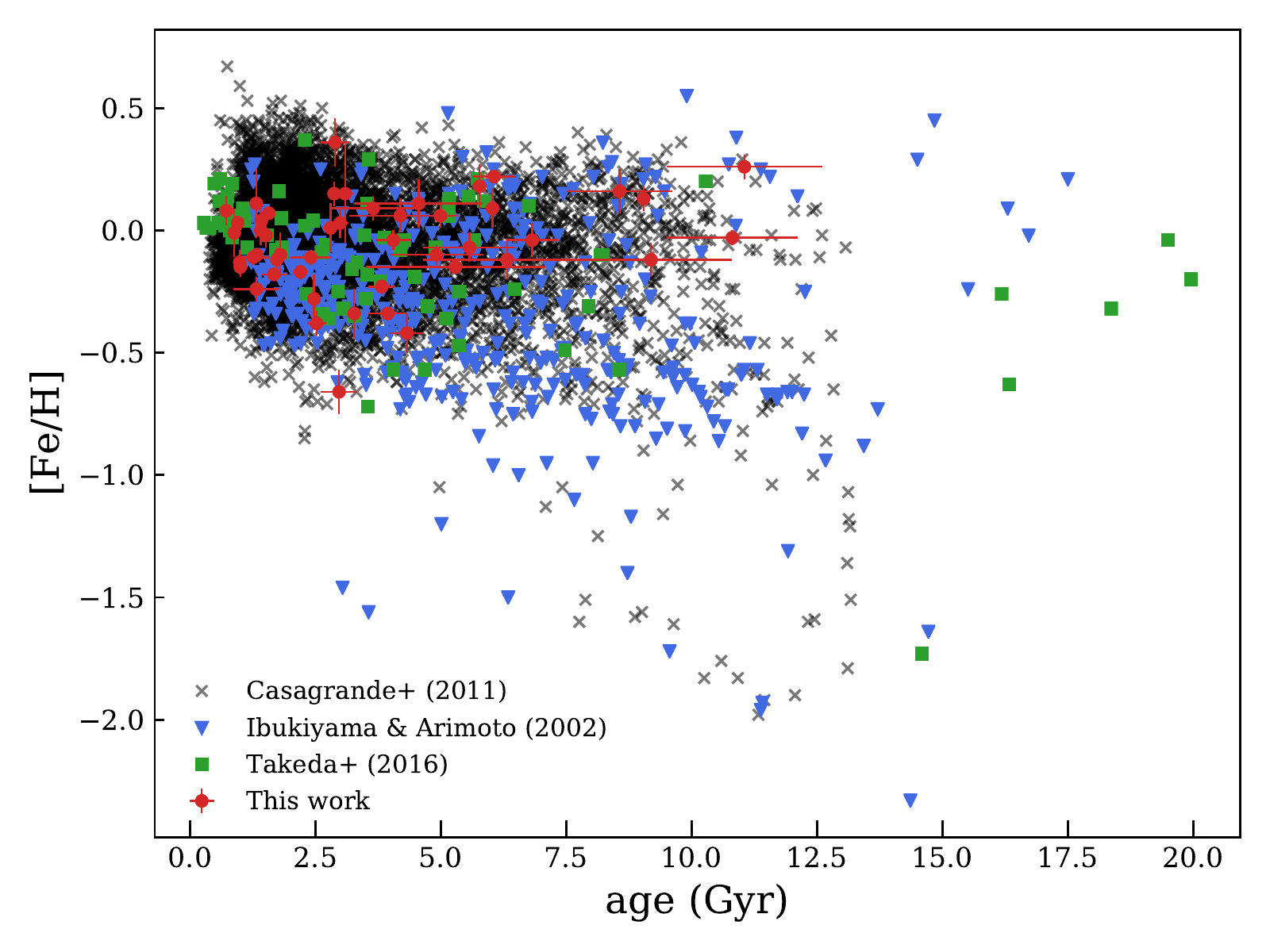}}
                \caption{Age-metallicity relation for our red clump stars compared to other Galactic works.}
                \label{figure_amr}
        \end{figure}

        We plotted our derived ages and metallicities in the age-metallicity diagram, and compared them with other studies for the solar neighbourhood. We first compared our derived ages with the work of \citet{Takeda_2016_04_0} for giant stars. We notice a very good agreement with their sample, as seen in Fig.~\ref{figure_amr}. In a wider context, our values for giants stars are also similar to the work for F-, G-, and K-type stars dwarfs \citep{Casagrande_2011_06_0,Ibukiyama_2002_11_0}. Our work supports the previous conclusions about the metallicity in our neighbourhood, that is, a little metallicity evolution in the past 10\,Gyr and a large scatter at all ages. Although the scatter of the relation seems to increase with age, the trend tends to be almost flat.
        
        %\section{Surface brightness-color relation for late-type stars}
        
        \section{Conclusion}
        \label{section_conclusion}
        
        We report accurate angular diameters measurements of nearby giant stars. Our sample includes a total of 48 stars for which the diameter is measured to better than 2.7\,\%. These observations were initially carried out to improve the calibration of the surface brightness-colour relation of late-type stars and to measure absolute stellar dimension of late-type eclipsing binaries to less than 1\,\%. We then used such systems to measure the most accurate Large Magellanic Cloud distance at a level of 1\,\% \citep{Pietrzynski_2018__0}.
        
        Combining our angular diameters measurements with Hipparcos and Gaia DR2 parallaxes and spectroscopic effective temperatures, we determined linear radii and absolute luminosities with an average accuracy of 3\,\% and 6\,\%, respectively. We also fitted \texttt{PARSEC} and \texttt{BaSTI} model isochrones to derive the age and mass of these giant stars. The added value of interferometry is that the constraint on the mass and age imposed by the $R-T_\mathrm{eff}$ plane is much tighter than using $L-T_\mathrm{eff}$ only. We found an overall good agreement between our estimated masses and literature values, while age estimates are rather scattered. Although we have accurate knowledge of the stellar angular diameters, our analysis still requires accurate determinations of the other stellar parameters such as the metallicity in order to be able to constrain different input physics and parameters from stellar evolution models.
        The stars of our sample will soon be observed by the TESS \citep{Ricker_2014_08_0} and PLATO \citep{Rauer_2014_11_0} satellites, that will provide detailed asteroseismic frequency spectra. Together with our high-precision interferometric angular diameters and Gaia distances, this will enable a more accurate determination of their physical parameters \citep[see e.g.][]{Kervella_2003_06_0,Thevenin_2005_06_0,Cunha_2007_11_0,Huber_2012_11_0}. This will also provide a stringent test of the asteroseismic scaling relations \citep[see e.g.][and reference therein]{Huber_2011_12_0,Gaulme_2016_12_0}.

        %--------------------ACKNOWLEDGEMENTS--------------------
        
        \begin{acknowledgements}

                The authors would like to thank all the people involved in the VLTI project.  A.~G. acknowledges support from FONDECYT grant 3130361. The authors acknowledge the support of the French Agence Nationale de la Recherche (ANR), under grant ANR-15-CE31-0012-01 (project Unlock-Cepheids). P.K., A.G., and W.G. acknowledge support of the French-Chilean exchange programme ECOS-Sud/CONICYT (C13U01). W.G., R.E.M. and G.P. gratefully acknowledge financial support for this work from the BASAL Centro de Astrofisica y Tecnologias Afines (CATA) PFB-06/2007. R.E.M. acknowledges grant VRID 218.016.004-1.0.W.G. also acknowledges financial support from the Millenium Institute of Astrophysics (MAS) of the Iniciativa Cientifica Milenio del Ministerio de Economia, Fomento y Turismo de Chile, project IC120009. We acknowledge financial support from the Programme National de Physique Stellaire (PNPS) of CNRS/INSU, France. The research leading to these results has received funding from the European Research Council (ERC) under the European Union’s Horizon 2020 research and innovation programme (grant agreement No. 695099). This work made use of the SIMBAD and VIZIER astrophysical database from CDS, Strasbourg, France and the bibliographic information from the NASA Astrophysics Data System. This research has made use of the Jean-Marie Mariotti Center \texttt{SearchCal} and \texttt{ASPRO} services, co-developed by FIZEAU and LAOG/IPAG, and of CDS Astronomical Databases SIMBAD and VIZIER. This work has made use of data from the European Space Agency (ESA) mission {\it Gaia} (\url{https://www.cosmos.esa.int/gaia}), processed by the {\it Gaia} Data Processing and Analysis Consortium (DPAC, \url{https://www.cosmos.esa.int/web/gaia/dpac/consortium}). Funding for the DPAC has been provided by national institutions, in particular the institutions participating in the {\it Gaia} Multilateral Agreement.
                
        \end{acknowledgements}
        
%--------------------BIBLIOGRAPHY--------------------

        \bibliographystyle{aa}   % if natbib is available
       \bibliography{bibliographie}
%        \bibliography{/Users/agallenn/Sciences/Articles/bibliographie}

%--------------------APPENDIX--------------------

\begin{appendix} %First appendix
        \section{Revised orbit of TZ~Fornacis}
        \label{appendix__new_orbit}
        
\begin{figure}
        \centering
        \resizebox{\hsize}{!}{\includegraphics[width = \linewidth]{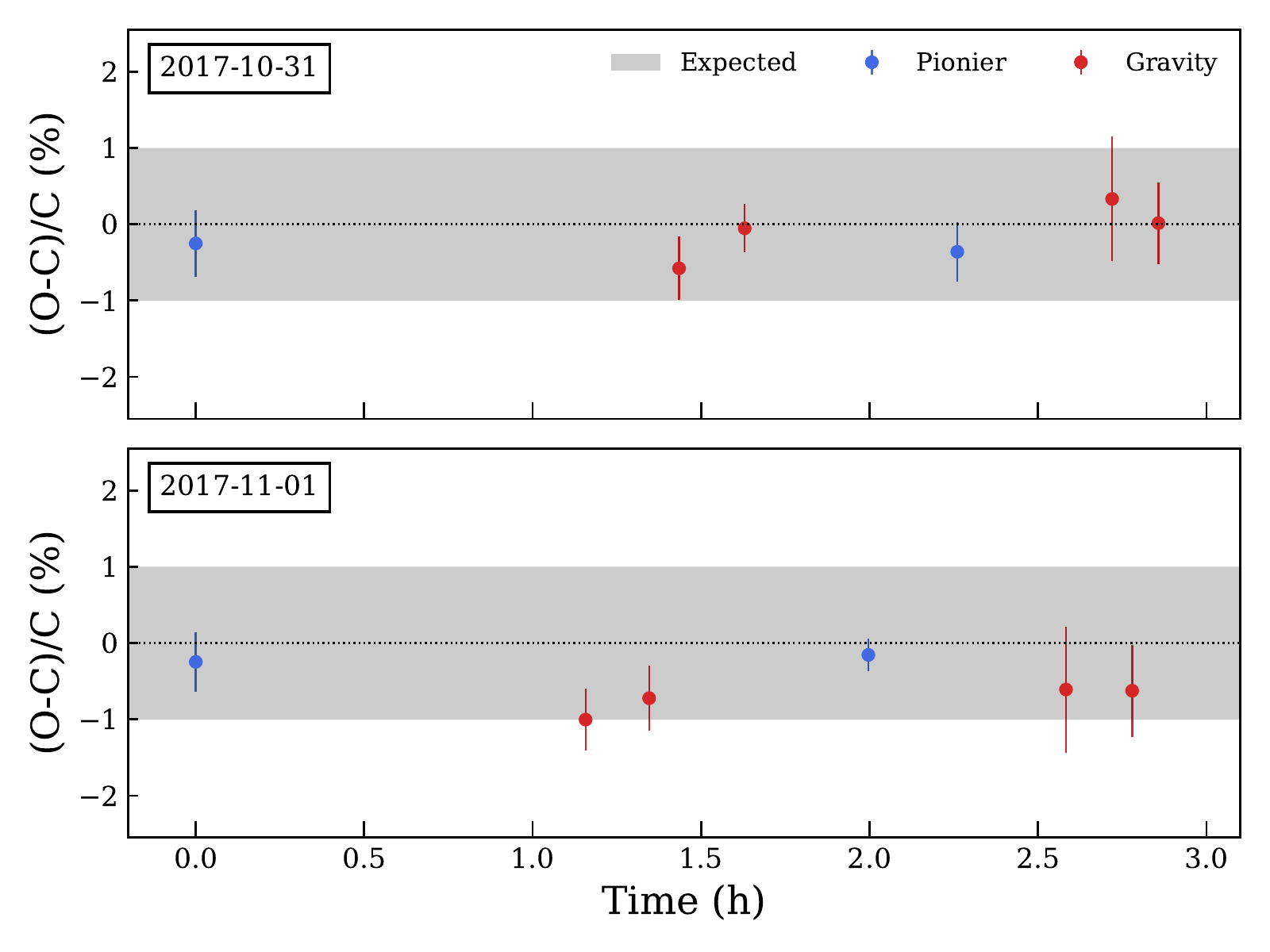}}
        \caption{Relative difference of the observed and newly calculated projected separations.}
        \label{figure_wave2}
\end{figure}

        Using the new interferometric measurements of this paper, we revised the orbital solutions of \citet{Gallenne_2016_02_0}  performing the exact same analysis, that is, simultaneously fitting the radial velocities of both components and the astrometry. The revised parameters are listed in Table~\ref{table__results}, which are in very good agreement (within $1\sigma$) with our previously determined values. Here, we have taken into account the systematic uncertainty from the wavelength calibration determined in this paper, reducing the uncertainty on the semi-major axis and the distance. The relative difference between the observed and new calculated projected separations are displayed in Fig.~\ref{figure_wave2}.
        
        \begin{table}[!ht]
                \centering
                \caption{Best-fit orbital elements and parameters.}
                \begin{tabular}{cc}
                        \hline
                        \hline
                        $P_\mathrm{orb}$ (days)                                                                 & $75.66691 \pm  0.00019$                                 \\
                        $T_\mathrm{p}$ (HJD)                                                                    &       2452599.29040    \\
                        $e$                                                                                                                                  &   $0.0000 \pm 0.0001$                        \\
                        $K_1$ ($\mathrm{km~s^{-1}}$)                                      &      $38.91 \pm 0.01$                                                \\
                        $K_2$ ($\mathrm{km~s^{-1}}$)                                      &      $40.88 \pm 0.01$                                                \\
                        $\gamma_1$      ($\mathrm{km~s^{-1}}$)                  &       $17.99 \pm 0.03$                                                       \\
                        $\gamma_2$      ($\mathrm{km~s^{-1}}$)                  &       $18.35 \pm 0.11$                                                       \\
                        $\omega$        ($\degr$)                                                                        &       $270.01 \pm 0.04$                                       \\
                        $\Omega$        ($\degr$)                                                                        &       $65.95 \pm 0.04$                        \\
                        $a$ (mas)                                                                                                                &        $2.990 \pm 0.011$          \\
                        $a$ (AU)                                                                                                                 &        $0.5565 \pm 0.0001$            \\
                        $i$ ($\degr$)                                                                                                   &       $85.71 \pm 0.04$                       \\
                        \hline
%                       $\theta_\mathrm{LD,1}$\tablefootmark{a}  &              $0.418 \pm 0.023$      &               $0.414 \pm      0.010$  \\
%                       $\theta_\mathrm{LD,2}$\tablefootmark{a}  &              $0.199 \pm 0.012$      &               $0.197 \pm      0.008$  \\
                        $M_1$ ($M_\odot$)                                                       &       $2.057 \pm 0.001$  \\
                        $M_2$ ($M_\odot$)                                                       &       $1.958 \pm 0.001$   \\
                        $d$ (pc)                                                                                        &  $186.1 \pm 0.7$    \\
                        $\pi$ (mas)                                                                                     &  $5.37 \pm 0.02$    \\
%                       $R_1$ ($R_\odot$)                                                       & $8.32 \pm 0.12$ & $8.28 \pm 0.22$ \\
%                       $R_2$ ($R_\odot$)                                                       & $3.96 \pm 0.09$ & $3.94 \pm 0.17$ \\
%                       $\log L_1/L_\odot$                                                      &       $1.59 \pm 0.04$       &       $1.57 \pm 0.02$ \\
%                       $\log L_2/L_\odot$                                                      &       $1.36 \pm 0.03$       &       $1.36 \pm 0.03$ \\
                        \hline          \end{tabular}
                \label{table__results}
        \end{table}

\end{appendix}

\end{document}